\documentclass[aps,prd,a4paper,showpacs,12pt]{article}
\usepackage[margin=1.0in]{geometry}
\usepackage{multicol}
\usepackage{flushend}
\usepackage[toc,page]{appendix}%%%%%%%%%%%%%%%%%%%%%%%%%%%%%%%%%%%%%%%%%%%%%%%%%%%%%%%%%%% New
\usepackage{subfigure}
\usepackage {graphicx}
\usepackage{color}
\usepackage{xcolor}
\usepackage{threeparttable}
\usepackage{amsfonts}
\usepackage{times}
\usepackage{setspace}
\usepackage{balance}
\usepackage{lastpage}
\usepackage{amsthm}
\usepackage[tbtags]{amsmath}
\usepackage{nomencl}
\usepackage{pstricks}
\usepackage{pstricks-add}
\usepackage{pst-plot,pstricks-add}

\usepackage{amsmath,amsfonts,amssymb,simplewick}
\usepackage{float}
\usepackage{amsmath,amssymb,amsfonts,epsfig,graphicx,euscript}%
\usepackage{hyperref} 
\hypersetup{bookmarks=true, bookmarksnumbered=true,linktoc=page, pdfstartview={FitH}, 
colorlinks=true, 
citecolor=red, 
filecolor=blue, 
linkcolor=blue, 
urlcolor=blue}
\usepackage{amsmath}% http://ctan.org/pkg/amsmath
\usepackage{amsfonts}% http://ctan.org/pkg/amsfonts
\usepackage{graphicx}
\usepackage{caption}
\usepackage{float}
\usepackage[all]{hypcap}
\numberwithin{equation}{section}
\usepackage{cite}
\usepackage{calc}
\newlength{\spacer}
\newsavebox{\mybox}

\newcommand{\bse}{\begin{subequations}}
\newcommand{\ese}{\end{subequations}}
\newcommand{\be}{\begin{equation}}
\newcommand{\ee}{\end{equation}}
\newcommand{\bea}{\begin{eqnarray}}
\newcommand{\eea}{\end{eqnarray}}
\newcommand{\ba}{\begin{array}}
\newcommand{\ea}{\end{array}}

\renewcommand{\thefootnote}{\fnsymbol{footnote}}
\begin{document}
%\hfill%
%\vbox{
%    \halign{#\hfil        \cr
%           IPM/P-2012/010\cr
%                     }}
%\vspace{1cm}
\begin{center}
{ \large{\textbf{The generation of baryon asymmetry and hypermagnetic field by the chiral vortical effect in the presence of sphalerons}}}
%{ \large{\textbf{Temperature Dependent Chiral Vortical Effect on Magnetogenesis and Baryogenesis}}} %\\
\vspace*{1.5cm}
%\vspace*{1.5cm}
\begin{center}
{\bf S. Abbaslu\footnote{s$_{-}$abbasluo@sbu.ac.ir}$^{1,2}$, A. Rezaei\footnote{amirh.rezaei@alumni.sbu.ac.ir}$^1$, S. Rostam Zadeh\footnote{sh$_{-}$rostamzadeh@ipm.ir}$^3$ and S. S. Gousheh\footnote{ss-gousheh@sbu.ac.ir}$^1$, }\\
\vspace*{0.5cm}
{\it{$^1$Department of Physics, Shahid Beheshti University, Tehran, Iran\\$^2$School of Physics, Institute for Research in Fundamental Sciences (IPM), PO Box 19395-5531, Tehran, Iran
		\\$^3$School of Particles and Accelerators, Institute for Research in Fundamental Sciences (IPM), P.O.Box 19395-5531, Tehran, Iran}}\\
\vspace*{1cm}
\end{center}	
\end{center}
\begin{center}
\today
\end{center}
%\vspace{.5cm}
%\bigskip

\renewcommand*{\thefootnote}{\arabic{footnote}}
\setcounter{footnote}{0}

%\begin{center}
\date{\today}
\textbf{Abstract:}
We show how the temperature-dependent chiral vortical effect can generate hypermagnetic fields and matter-antimatter asymmetries, in the symmetric phase of the early Universe, in the temperature range $100\mbox{GeV} \le T\le 10\mbox{TeV}$, even in the presence of the weak sphaleron processes. We take into account all  perturbative chirality-flip processes, as well as the nonperturbative Abelian and non-Abelian anomalous effects for all three generations. Using the constraints and conservation laws in the plasma, we reduce the number of required evolution equations. We also simplify the anomalous transport coefficients, accordingly. We consider both monochromatic and continuous spectra for the hypermagnetic and velocity fields to solve the anomalous magnetohydrodynamics equations. We then show that overlapping  small transient fluctuations in the temperature of some matter degrees of freedom and vorticity of the plasma can generate a chiral vortical current, resulting in the generation of strong hypermagnetic fields and matter-antimatter asymmetries, all starting from zero initial values. We obtain the baryon asymmetry  $\eta_{B}\simeq 5\times10^{-10}$ and a positive helicity hypermagnetic field with amplitude $B_{Y}(x)\simeq10^{19}G$, at the onset of the electroweak phase transition.
Although the sphaleron processes tend to washout the generated $(\rm B +L)$ asymmetry, the anomalous processes prevail and the baryogenesis and leptogenesis occur without $(\rm B-L)$ violation.

\section{Introduction}  
One of the open problems in modern cosmology and particle physics is the baryon asymmetry of the Universe. In the current cosmological models it is assumed that matter and antimatter are created equally at the time of the Big Bang. However, observations indicate that our observable Universe, out to the Hubble size, is made almost entirely of matter. The measured amplitude of the baryon asymmetry is of the order of $\eta_{\mathrm{B}}\sim 10^{-10}$ \cite{{bas1},{bas2},{bas3}}. Although many studies have been conducted in this regard, the origin of this asymmetry is still under debate \cite{Giovannini-1997eg,Giovannini2000,Giovannini-2013oga,Giovannini-2015aea,Giovannini-2016whv,basy1,basy2,Joyce-1997uy,basy5,basy6,basy8,h1,sh3,n4,n5,n1}. In all CPT invariant models, the following three necessary Sakharov conditions should be satisfied for producing the baryon asymmetry \cite{Sakharov1}:
baryon number violation,  C and CP violation\footnote{Under the C and CP symmetries, the creation rates of matter and antimatter will be the same.}, and departure from thermal equilibrium.\footnote{In thermal equilibrium, the number of particles for each species is equal to number of antiparticles.}

The first condition is satisfied in the symmetric phase through the anomalies \cite{a1,a2,a3}. In fact, conservation of the fermionic matter currents is no longer valid in the presence of the $\textrm{SU}_{\textrm{L}}(2)$ and $\textrm{U}_{\textrm{Y}}(1)$ gauge fields \cite{basy6}.\footnote{ The nonperturbative high temperature effects associated with the $\textrm{SU}(3)$ anomaly, known as the strong sphaleron processes, change the chiralities of the quarks but respect the quark number conservation \cite{qn1}. The reaction rate is $\Gamma_{s}\simeq100 \alpha_{s}^{5}T$ \cite{ssp}, where $\alpha_{s}$ is the $\textrm{SU}(3)$ fine structure constant.}
The corresponding anomalies violate the baryon and lepton numbers separately, but preserve $B-L$, where $B$ $(L)$ denotes the total baryon (lepton) number \cite{su1}. 
The nonperturbative high temperature effects associated with the $\textrm{SU}_{\textrm{L}}(2)$ are known as the weak sphaleron processes with the reaction rate $\Gamma_{w}\simeq25 \alpha_{w}^{5}T$ \cite{wsp1,wsp2}, where $\alpha_{w}$ is the $\textrm{SU}_{\textrm{L}}(2)$ fine structure constant. The weak sphaleron processes violate the baryon and lepton numbers through converting, for example, nine quarks (antiquarks)  into three antileptons (leptons) or vice versa. 

The Abelian $\textrm{U}_{\textrm{Y}}(1)$ anomaly violates the baryon and lepton numbers separately, due to the chiral coupling of the $\textrm{U}_{\textrm{Y}}(1)$ gauge field to the fermions \cite{Giovannini-1997eg,Joyce-1997uy,basy5,basy6}.  The Abelian anomaly does not have sphaleron-like processes, and fermion number violation occurs only due to the time variation of the hypermagnetic field helicity, which has been widely investigated in the literature \cite{Giovannini-1997eg,Giovannini-2015aea,n1,h1,Giovannini-2013oga,Giovannini2000,Giovannini-2016whv, Joyce-1997uy,basy6,basy8,n5,n4}. Moreover, the Abelian $\textrm{U}_{\textrm{Y}}(1)$ gauge field, in contrast to the non-Abelian ones, remains  massless at finite temperatures and can produce long-range hypermagnetic fields \cite{hf1,hf2}, which are transformed to the ordinary Maxwellian magnetic fields during the electroweak phase transition (EWPT).

The existence of the long-range magnetic fields, ubiquitous in all observable Universe from stars to galaxies and intergalactic medium, is another open problem in particle physics and cosmology \cite{abc1,abc2,abc3}. The reported strength of the coherent magnetic fields in the Milky Way and in the intergalactic medium is of the order of $10^{-6}\mbox{G}$ and $10^{-15}\mbox{G}$, respectively \cite{abc1,magnetic2,Ando:2010rb,abc2,magnetic5}.
Various models and mechanisms have been proposed to explain the origin of these magnetic fields, the most 
widely-studied of which are the astrophysical and the cosmological ones \cite{abc4,abc5,abc6,abc7}. The astrophysical mechanisms are local models applicable during and after the structure formation \cite{abc3},
while the cosmological mechanisms are global models applicable even before the EWPT \cite{abc4}. In this paper we use 
a vorticity-based model which is in the latter category \cite{s1,Giovannini-1997eg,Giovannini-2015aea,Giovannini-2013oga}. It has been shown that in a chiral vortical plasma at high temperatures ($T\ge100 \rm GeV$), the hypermagnetic fields can be generated without any initial seed \cite{s1,Giovannini-1997eg,Giovannini-2015aea,Giovannini-2013oga,s3}. 

One of the most important effects that plays a significant role in magnetogenesis and baryogenesis is the chiral vortical effect (CVE) \cite{s1,Giovannini-1997eg,Giovannini-2015aea,Giovannini-2013oga,s3}. It was discovered by Vilenkin, when he showed that a rotating black hole can produce a chiral neutrino matter current parallel to the vorticity\cite{Vilenkin}. Afterwards, it was shown that in a chiral vortical plasma, a similar chiral current, $ \vec{j}_{\mathrm{cv}R,L}=\pm\left[\frac{1}{8\pi^2}\mu_{R,L}^{2}\right]\vec{\omega}$  exists for any chiral fermion, where $+ (-)$ is for the right-handed (left-handed) fermion, $\mu_{R,L}$ is its chiral chemical potential, and $\vec\omega$ is the vorticity of the plasma \cite{v1,v2}. A more complete form of this current is $ \vec{j}_{\mathrm{cv}R,L}=\pm\left[\frac{1}{8\pi^2}\mu_{R,L}^{2}+\frac{1}{24}T^2\right]\vec{\omega}$ \cite{Vilenkin,v8}, and therefore the CVE  can be activated either through the chemical potential or through the temperature-dependent term in the bracket. The former can produce an initial seed for the hypermagnetic field \cite{s1}, while the latter can generate and amplify the hypermagnetic field, and therefore matter-antimatter asymmetries, all starting from zero initial values \cite{Temp1,s3,s1,Giovannini-2013oga,Giovannini-2015aea}. The latter mechanism is possible only if the out-of-equilibrium conditions include a temperature difference between the matter degrees of freedom. These are usually considered to be in the form of localized fluctuations, but the corresponding global fluctuations in the early Universe have also been considered \cite{Gangopadhyay-2009dv} \footnote{In fact, these out-of-equilibrium conditions in a simple electromagnetic plasma are usually considered to be in the latter form, {\it i.e.}, the electrons and ions being at different global temperatures  \cite{61-1,Maximizing}.}.

In a magnetized chiral plasma, another non-dissipative matter current appears parallel to the magnetic field. This effect is known as the chiral magnetic effect (CME) \cite{kh1,kh2,kh3,kh4}. The matter chiral magnetic current for a chiral fermion in the broken phase is obtained as $ \vec{j}_{\mathrm{cm}R,L}=\pm\frac{Q_{R,L}}{4\pi^2}\mu_{R,L}\vec{B}$, where $Q_{R,L}$ is the electric charge of the chiral fermion \cite{v8,kh4,wang1}. More importantly, in the symmetric phase
the CME and the $\textrm{U}_{\textrm{Y}}(1)$ Abelian anomaly interconnect the hypermagnetic helicity and fermion number densities \cite{a3,Joyce-1997uy,Giovannini-1997eg}. 

Many studies have investigated the role of the anomalous transport effects, {\it i.e.}, the CVE and the CME, in the evolution of the hypermagnetic fields and the matter-antimatter asymmetries \cite{Giovannini-1997eg,Giovannini2000,Giovannini-2013oga,Giovannini-2015aea,Giovannini-2016whv,Joyce-1997uy,basy5,basy6,basy8,h1,sh3,n4,n5,n1}. An initially small vorticity and a large matter asymmetry stored in the form of right-handed electrons at $10$ TeV, have been considered in Ref.\ \cite{s2}, leading to the production of the hypermagnetic field and the baryon asymmetry in the temperature range, $100\mbox{ GeV} \le T\le 10\mbox{ TeV}$.
Furthermore, the effects of overlapping transient fluctuations in the temperature of some matter degrees of freedom and vorticity of the plasma have been studied in Ref.\ \cite{s3}, resulting  in the generation of the hypermagnetic field and the matter-antimatter asymmetries, all starting from zero initial values.
In both of these studies, only the effects of the baryons and the first-generation leptons have been considered in the anomalous transport coefficients, and the contributions of the second- and third-generation leptons, and the Higgs boson have been ignored. Moreover, fast interactions such as the Yukawa processes of the second- and third-generation leptons, and also the weak sphaleron processes, which can wash out the asymmetries of the baryons and leptons, have been neglected \cite{n4,jdvm,basy8}. In this paper we generalize our previous work \cite{s3} by taking into account all of the neglected components and processes mentioned above. Furthermore, in the final part of this work, we extend the monochromatic Chern-Simons configurations of the hypermagnetic and velocity fields to continuous spectra.

As is well known, the weak sphaleron processes couple to the left-handed fermions and affect their evolution directly. However, the weak sphalerons cannot affect the right-handed fermions unless they have fast Yukawa interactions with the left-handed ones. Therefore, when the weak sphaleron processes are in thermal equilibrium, the Yukawa interactions play an important role in the production of baryon and lepton asymmetries. More generally, our Universe in its early stages was a hot plasma, consisting of all fermions and bosons, having many perturbative and nonperturbative interactions with one another. Below we shall show that for every fast process there is an associated equilibrium condition with a corresponding conserved physical quantity, which can be used as a constraint to reduce the number of variables in the evolution equations. In the temperature range of our interest, many of these processes are extremely fast, each leading to a potential constraint. Therefore, in order to have a more realistic model, we will consider these constraints along with the absolute conservation laws such as the hypercharge neutrality condition, the Abelian and non-Abelian anomalous effects, the fermionic Yukawa interactions, and also the contribution of all fermionic and bosonic chemical potentials to the anomalous transport coefficients. 
We should mention that one cannot properly take into account these constraints and conservation laws without including the chemical potentials of all particles, {\it i.e.} leptons, quarks and the Higgs. With the proper inclusion of these constraints and conservation laws, as well as the sphaleron processes, some of our results change considerably, as compared to our previous work \cite{s3}, which we shall point out in Sec.~\ref{x3}.

This paper is organized as follows: In Sec.\ \ref{x2} we consider the perturbative chirality-flip processes, the fast sphaleron processes, and the relevant constraints along with the conservation laws to obtain the relations between the chemical potentials of the particles in the symmetric phase of the early Universe. In Sec.\ \ref{x1} we present the evolution equations for the fermionic asymmetries and the hypermagnetic field. In Sec.\ \ref{x3} we numerically solve the set of coupled
differential equations obtained in Sec.\ \ref{x1} for the hypermagnetic
field and the fermionic asymmetries all starting from zero initial values.
In Sec.\ \ref{x4} we discuss our results. In Appendix.\ \ref{xx} we present the anomaly equations for the matter currents, including $\textrm{U}_\textrm{Y}$(1), $\textrm{SU}_\textrm{L}(2)$ and $\textrm{SU}(3)$ anomalies, as well as the conserved $\textrm{U}_\textrm{Y}$(1) hypercharge current. In Appendix.\  \ref{x1b} we present the Anomalous Magnetohydrodynamic (AMHD) equations, taking the CVE and the CME into account, by considering the monochromatic Chern-Simons configuration for the velocity and hypermagnetic fields. In Appendix.\  \ref{app-c} we extend the AMHD equations obtained in Appendix.\  \ref{x1b}  by considering continuous spectra for the hypermagnetic and velocity fields. In the following we use the natural units, in which $\hbar=c=1$, and also the Friedmann-Robertson-Walker (FRW) metric $ds^{2}=dt^{2}-R^{2}(t)\delta_{ij}dx^{i}dx^{j}$, where $t$ is the physical time, and $x^{i}$s are the comoving coordinates.

\section{Conservation Laws and Equilibrium Conditions}\label{x2}

In this section, we use the constraints derived from fast processes in the electroweak plasma, along with the conservation laws, to obtain simple relations between the chemical potentials in the temperature range, $100\mbox{GeV} \le T\le 10\mbox{TeV}$. To do this, let us denote the chemical potential of the Higgs boson by $\mu_{{\varPhi}^+}=\mu_{{\varPhi}^0}\equiv\mu_{0}$, the left-handed quarks with different colors by $\mu_{u_L^i}=\mu_{d_L^i}\equiv\mu_{Q^{i}}$, the right-handed up (down) quarks with different colors by $\mu_{u_R^i}$ ($\mu_{d_R^i} $) and the right-handed (left-handed) leptons by $\mu_{e_R^i}$ ($ \mu_{e_L^i}=\mu_{\nu_L^i} $), where `$i$' is the generation index.\footnote{Note that, in the temperature range of interest, fast gauge interactions maintain the equality of the asymmetries carried by different components within a given multiplet.}

As is well known, the non-Abelian $\rm SU(3)$ and $\rm SU_{L}(2)$ gauge theories have degenerate vacua, each vacuum labeled with a different integer Chern-Simons number. As mentioned before, the rate of the strong sphaleron interactions is high and these processes are in thermal equilibrium ($ \Gamma_{s}>H$) below the temperature $T_{s}\simeq10^{15} \mbox{GeV}$, where $H= \sqrt{\frac{4\pi^3 g^{*}}{45}}\frac{T^2}{M_{Pl}}$ is the Hubble parameter, $g^*=106.75$ is the effective number of relativistic degrees of freedom, and $M_{Pl}=1.22\times10^{19} \mbox{GeV}$ is the Planck mass. The strong sphaleron processes only change the chirality of the quarks by simultaneously converting all left-handed quarks, one from each species and color, to right-handed quarks, and vice versa, {\it i.e.}, $\sum_{i} \left(u_L^i+d_L^i\right) \leftrightarrow \sum_{i} \left(u_R^i+d_R^i\right) $. Therefore, in thermal equilibrium, they provide a further constraint on the chemical potentials of the quarks of all generations as \cite{h1}
\begin{equation}\label{eqss}
c_{s}\equiv\sum_{i} [2\mu_{Q^i}-\mu_{d_R^i}-\mu_{u_R^i}]=0.
\end{equation} 
Furthermore, due to the flavor mixing in the quark sector, all up or down quarks belonging to different generations with distinct handedness have the same chemical potential, i.e., $\mu_{u_R^i}=\mu_{u_{R}}$,  $\mu_{d_R^i}=\mu_{d_{R}}$, and $\mu_{Q^{i}}=\mu_{Q}$. Therefore, Eq.\ (\ref{eqss}), which refers to the chemical equilibrium of the strong sphaleron process reduces to the simple form  \cite{sh3,h1} 
\begin{equation}\label{eqss1}
2\mu_{Q}-\mu_{d_{R}}-\mu_{u_{R}}=0.
\end{equation}

Another sphaleron process is the weak sphaleron process, which in thermal equilibrium might wash out the baryon and lepton asymmetries by simultaneously converting color singlet quarks to antileptons, and vice versa, {\it i.e.},  $\sum_{i} \left(3u_L^i++3d_L^i+e_L^i+\nu_L^i\right) \leftrightarrow 0 $. The rate of these sphaleron processes above the electroweak scale is estimated by the numerical simulations to be $\Gamma_{w} \simeq 25\alpha_{w}^5 T$, where $\alpha_{w}=g^{2}/4\pi$ is the $\textrm{SU}_{\textrm{L}}(2)$ fine structure constant \cite{wsp1,wsp2}. These processes are in thermal equilibrium ($ \Gamma_{w}>H$) below the temperature $T_{w}\simeq10^{12} \mbox{GeV}$, providing a further constraint on the chemical potentials of left-handed quarks and leptons of all generations as \cite{h1}
\begin{equation}\label{eq.sph}
c_{w}\equiv\sum_{i} N_w[3\mu_{Q^i}+\mu_{e_L^i}]=2(9\mu_{Q}+\mu_{e_L}+\mu_{\mu_L}+\mu_{\tau_L})=0,
\end{equation}
 where $ N_w=2$ is the rank of $\rm SU(2)$.
In contrast to the vacuum structure of the $\rm SU(3)$ and $\rm SU_{L}(2)$ gauge theories, the vacuum structure of the $\textrm{U}_{\textrm{Y}}(1)$ gauge theory is trivial. However, as we shall see, there can be a nontrivial hypermagnetic field with a time-varying helicity in the symmetric phase. The time evolution of the hypermagnetic helicity violates the fermion numbers, in accordance with the anomaly equations.\footnote{Note that the effect of this helicity is different in the broken phase, where it changes only the chirality of the fermions.}

There are also the perturbative chirality-flip processes operating on the quarks and the leptons with the rates $\Gamma_{i}\simeq 10^{-2}h_{i}^{2}T/8\pi $, where the Yukawa couplings $h_i$ are given as \cite{Davidson-1994gn,n5,Kawamura-2018cpu}
\begin{equation}
\begin{split}
&h_{e}\simeq2.8\times10^{-6},\qquad h_{\mu}\simeq5.8\times10^{-4},\qquad h_{\tau}\simeq10^{-2},\\&
h_{u}\simeq1.1\times10^{-5},\qquad h_{c}\simeq7.1\times10^{-3},\qquad h_{t}\simeq0.94,\\&
h_d\simeq2.7\times10^{-5},\qquad h_s\simeq 5.5\times10^{-4},\qquad h_b \simeq2.4\times10^{-2}.
\end{split}
\end{equation}
The Yukawa interactions of the electron, muon, and tau are in thermal equilibrium below the temperatures 
$T_{e,\text{Yuk}} \approx 5\times 10^4 \, \text{GeV}$, 
$T_{\mu,\text{Yuk}} \approx 10^9 \, \text{GeV}$, and 
$T_{\tau,\text{Yuk}} \approx 5\times 10^{11} \, \text{GeV}$, respectively. Therefore, in the temperature range $\mbox{T}_{EW}\le T\le10\mbox{TeV}$, the weak and strong sphaleron processes, and the Yukawa interactions of all quarks and leptons are in chemical equilibrium. However, it has been shown that in the presence of the strong hypermagnetic fields, these processes can fall out of chemical equilibrium. The amount of falling out of equilibrium depends on the rate of the process, i.e., the slower the process, the more intensely it falls out of chemical  equilibrium\cite{sh3,sphaleron1}.  The rate of the weak sphaleron processes is greater than those of all lepton chirality-flip processes. Therefore, in this work, where the strong hypermagnetic fields are also present, we will let the chirality-flip processes of all leptons to be out of chemical equilibrium but assume that the sphaleron processes and the chirality-flip processes of all quarks are still nearly in chemical equilibrium.\footnote{Although some of the quark Yukawa couplings are small compared to leptons, all quark-Yukawa interactions remain in thermal equilibrium due to the combined effects of strong sphaleron processes and the quark mixing, below the temperature which is relevant for the top quark, {\it i.e.}, $T_{t,\mathrm{Yuk}}\simeq10^{12}\mbox{GeV}$.}
Therefore, the following constraints are obtained for the chemical potentials of the quarks: \cite{h1}
\begin{equation}
\begin{split}\label{eq.mu11}
&\mu_{u_R^i}-\mu_{Q^{i}}=\mu_{u_{R}}-\mu_{Q}=\mu_0,\quad \quad i=1,2,3,\\&
\mu_{d_R^i}-\mu_{Q^{i}}=\mu_{d_{R}}-\mu_{Q}=-\mu_0,\quad \quad i=1,2,3.\\&
%\mu_{R_{i}}-\mu_{L_{i}}=-\mu_0\quad \quad i=3
\end{split}
\end{equation}
Using Eqs.\ (\ref{eq.mu11}), the total baryonic chemical potential can be obtained as
\begin{equation}\label{eq.mu2}
\begin{split}
\mu_{B}=\sum_{i=1}^{n_{G}}\left[\mu_{d_{R}^i}+\mu_{u_{R}^i}+{ N_{w}}\mu_{Q^i}\right]=12\mu_Q.
\end{split}
\end{equation}
Using Eqs.\ (\ref{eq.jy}), and (\ref{eq.mu11}), the hypercharge neutrality condition can also be obtained as
\begin{equation}\label{eq.mu31}
Q_Y=6\mu_Q-\mu_{e_L}-\mu_{e_R}-\mu_{\mu_R}-\mu_{\mu_L}-\mu_{\tau_R}-\mu_{\tau_L}+11\mu_{0}=0.
\end{equation}
Moreover, Eqs.\ (\ref{eq.as1})  lead to the following three $B-L$ conservation laws for three fermion generations: 
\begin{equation}\label{eta11}
\begin{split}
&\frac{1}{T}\left[\frac{\mu_{B}}{3}-\mu_{e_R}-2\mu_{e_L}\right]=c_1,\\&
\frac{1}{T}\left[\frac{\mu_{B}}{3}-\mu_{\mu_R}-2\mu_{\mu_L}\right]=c_2,\\&
\frac{1}{T}\left[\frac{\mu_{B}}{3}-\mu_{\tau_R}-2\mu_{\tau_L}\right]=c_3.\\&
\end{split}
\end{equation}

Since in our scenario all initial asymmetries are zero,  the above constant values $c_i$, $i=1,2,3$, are set to zero.
The use of the constraints given by equations (\ref{eq.sph}), (\ref{eq.mu11}), (\ref{eq.mu31}), and (\ref{eta11}), along with Eq.\ (\ref{eq.mu2}), leads to the minimum number of required dynamical equations for obtaining the evolution of all chemical potentials. We choose to solve the evolution equations for $\mu_{e_R}$, $\mu_{\mu_R}$, and $\mu_{\tau_R}$, and then obtain the evolution of all other chemical potentials via the relations 
\begin{equation}\label{mus}
\begin{split}
& \mu_{Q}=\frac{\mu_{e_R}+\mu_{\mu_R}+\mu_{\tau_R}}{30},\qquad\qquad\qquad\mu_{0}=\frac{\mu_{e_R}+\mu_{\mu_R}+\mu_{\tau_R}}{22},\\&
\mu_{e_L}=\frac{-13\mu_{e_R}+2\mu_{\mu_R}+2\mu_{\tau_R}}{30},\qquad\qquad\mu_{\mu_L}=\frac{2\mu_{e_R}-13\mu_{\mu_R}+2\mu_{\tau_R}}{30},\\&
\mu_{\tau_L}=\frac{2\mu_{e_R}+2\mu_{\mu_R}-13\mu_{\tau_R}}{30}.
\end{split}
\end{equation}

In the next section, these relations will be used for obtaining the hypercharge chiral magnetic and chiral vortical coefficients.

\section{The evolution equations}\label{x1}

In this section, we present the evolution equations for the hypermagnetic field and matter-antimatter asymmetries, taking into account the CME, the CVE, and all of the conditions obtained in Sec.\ \ref{x2}. Here, we present the AMHD equations for the monochromatic Chern-Simons wave configuration of the hypermagnetic and velocity fields. The extended equations for continuous spectra of the fields are presented in Appendix~\ref{app-c}.
Using the anomaly equation (\ref{er}) and taking the relevant chirality-flip processes into account, we obtain the evolution equations for the asymmetries of the right-handed electron, muon, and tau as\footnote{High temperature of the early Universe plasma and low-velocity limit, imply that $j^{0}_\mathrm{r}\simeq (n_\mathrm{r}-\bar{n}_\mathrm{r})$, to a very good approximation \cite{s1 }.} \cite{n5}
\begin{equation}\label{as1}
\begin{split}
&\frac{d\eta_{{e}_{R}}}{dt}=\frac{g'^{2}}{4\pi^{2} s}\langle\vec{E}_{Y}.\vec{B}_{Y}\rangle+\Gamma_{e}\left(\eta_{e_{L}}-\eta_{e_{R}}-\frac{\eta_0}{2}\right),\\&
\frac{d\eta_{{\mu}_{R}}}{dt}=\frac{g'^{2}}{4\pi^{2} s}\langle\vec{E}_{Y}.\vec{B}_{Y}\rangle+\Gamma_{\mu}\left(\eta_{\mu_{L}}-\eta_{\mu_{R}}-\frac{\eta_0}{2}\right),\\&
\frac{d\eta_{{\tau}_{R}}}{dt}=\frac{g'^{2}}{4\pi^{2} s}\langle\vec{E}_{Y}.\vec{B}_{Y}\rangle+\Gamma_{\tau}\left(\eta_{\tau_{L}}-\eta_{\tau_{R}}-\frac{\eta_0}{2}\right),
\end{split}
\end{equation}
where we have used the relation $\mu=(6s/cT^2)\eta$, with $c=1$ for the fermions and $c=2$ for the bosons. Here $s=2\pi^{2}g^{*}T^3/45$ is the entropy density, and $\Gamma_i\simeq10^{-2}h_{i}^{2}T/8\pi=\Gamma_{i}^{0}/(\sqrt{x}t_{EW})$ are the rates of lepton Yukawa interactions, with $\Gamma_{e}^{0}=11.38$, $\Gamma_{\mu}^{0}=4.88\times10^{5}$, and $\Gamma_{\tau}^{0}=1.45\times10^{8}$ \cite{n5}. Here $x=t/t_\mathrm{EW}=\left(T_\mathrm{EW}/T\right)^{2}$ is given by the Friedmann law, $t_\mathrm{EW}=M_{0}/2T_\mathrm{EW}^{2}$, $M_{0}=M_\mathrm{Pl}/1.66\sqrt{g^{*}}$ is the reduced Planck mass, and $T_\mathrm{EW}$ is temperature of the EWPT. Equation (\ref{as1}) shows that the differences in the evolution equations for asymmetries of right-handed leptons are only due to their different chirality-flip rates. Using the expressions for the chiral magnetic coefficient $c_{\rm B}$, the chiral vortical coefficient $c_{\rm v}$, and the hyperelectric field, given by Eqs.\  (\ref{eq27}), (\ref{eq28}), and (\ref{eq16}), and also the helical configurations for the hypermagnetic and vorticity fields introduced in Eqs.\ (\ref{eq10}) and (\ref{eq11}), respectively, we obtain
\begin{equation}\label{eq44}
\begin{split}
\langle\vec{E}_{Y}.\vec{B}_{Y}\rangle=&\frac{B_{Y}^{2}(t)}{100} \left[\frac{k^{\prime}}{T}-\left(\frac{6sg'^{2}}{4\pi^{2}T^3}\right)\eta_{T}\right]\\&-\left[\frac{g'}{24}\beta(T)+\left(\frac{36s^2g'}{2\pi^{2}T^6}\right)\Delta\eta^{2}\right]\frac{k^{\prime}T }{100}\langle\vec{v}(t).\vec{B}_{Y}(t)\rangle,
\end{split}
\end{equation}
where $k^{\prime}=k/R=kT$ and
\begin{equation}\label{eq44as}
\begin{split}
& \Delta\eta^{2}=\eta_{e}\eta_{e}^{5}+\eta_{\mu}\eta_{\mu}^{5}+\eta_{\tau}\eta_{\tau}^{5}-\frac{3}{16}\eta_{0}\left(\eta_{B}+\eta_{0}\right),\\&
\eta_{T}=\frac{T^2}{6s}(c_{e, \mathrm{Yuk}}+c_{\mu, \mathrm{Yuk}}+c_{\tau, \mathrm{Yuk}})=\eta_{e, \mathrm{Yuk}}+\eta_{\mu, \mathrm{Yuk}}+\eta_{\tau,\mathrm{Yuk}},\\&
\eta_{e^i,\mathrm{Yuk}}=\eta_{e_R^i}-\eta_{e_L^i}+\frac{1}{2}\eta_{0}\qquad \text{for}\quad  i=e, \mu, \tau.
\end{split}
\end{equation}
Using the relation $1\mbox{Gauss}\simeq2\times10^{-20}  \mbox{GeV}^{2}$, and recalling $x=\left(t/t_\mathrm{EW}\right)=\left(T_\mathrm{EW}/T\right)^{2}$, we obtain the complete set of evolution equations for  $\eta_{e_R}$, $\eta_{\mu_R}$, $\eta_{\tau_R}$, and the amplitude of the hypermagnetic field as 
\begin{equation}\label{eq47}
\begin{split}
\frac{d \eta_{e_R}(x)}{dx}=&\left[C_{1}\left(\frac{k}{10^{-7}}\right)-C_{2} \eta_{T}(x)
\right]\left(\frac{B_{Y}(x)}{10^{20}G}\right)^{2}x^{3/2}\\&-\left[C_{3}\beta(x)+C_{4} \Delta \eta^{2}(x)\right]v(x)\left(\frac{k}{10^{-7}}\right)\left(\frac{B_{Y}(x)}{10^{20}G}\right)\sqrt{x}-\frac{\Gamma_{e}^{0}}{\sqrt{x}}\eta_{e,\mathrm{Yuk}}(x),
\end{split}
\end{equation}
% \textcolor{blue}{} 
\begin{equation}\label{eq54}
\begin{split}
\frac{d\eta_{\mu_{R}}(x)}{dx}=&\left[C_{1}\left(\frac{k}{10^{-7}}\right)-C_{2}\eta_{T}(x)
\right]\left(\frac{B_{Y}(x)}{10^{20}G}\right)^{2}x^{3/2}\\&-\left[C_{3}\beta(x)+C_{4} \Delta \eta^{2}(x)\right]v(x)\left(\frac{k}{10^{-7}}\right)\left(\frac{B_{Y}(x)}{10^{20}G}\right)\sqrt{x}-\frac{\Gamma_{\mu}^{0}}{\sqrt{x}}\eta_{\mu,\mathrm{Yuk}}(x),
\end{split}
\end{equation}
\begin{equation}\label{eq54-h}
\begin{split}
\frac{d\eta_{\tau_{R}}(x)}{dx}=&\left[C_{1}\left(\frac{k}{10^{-7}}\right)-C_{2}\eta_{T}(x)
\right]\left(\frac{B_{Y}(x)}{10^{20}G}\right)^{2}x^{3/2}\\&-\left[C_{3}\beta(x)+C_{4} \Delta \eta^{2}(x)\right]v(x)\left(\frac{k}{10^{-7}}\right)\left(\frac{B_{Y}(x)}{10^{20}G}\right)\sqrt{x}-\frac{\Gamma_{\tau}^{0}}{\sqrt{x}}\eta_{\tau, \mathrm{Yuk}}(x),
\end{split}
\end{equation}
\begin{equation}\label{eq49} 
\begin{split}
\frac{dB_{Y}(x)}{dx}=&\frac{1}{\sqrt{x}}\left[-C_{5}\left(\frac{k}{10^{-7}}\right)^2+ C_{6}\left(\frac{k}{10^{-7}}\right)\eta_{T}(x)
\right]B_{Y}(x)-\frac{1}{x}B_{Y}(x)\\&+\left(\frac{k}{10^{-7}}\right)^2\left[C_{7}\beta(x)+C_{8}\Delta \eta^{2}(x)\right]\frac{v(x)}{x^{3/2}} ,
\end{split}
\end{equation}
where $\alpha_{Y}=g'^{2}/4\pi\simeq0.01$   
and the coefficients $C_{i}, i=1,...,8$ are given by 
\begin{equation}\label{eq51}
\begin{split}
&
C_{1}=9.6\times10^{-4}\alpha_{Y},\\&
C_{2}=865688 \alpha_{Y}^{2},\\&
C_{3}=0.71488\alpha_{Y}^{3/2},\\&
C_{4}= 68610.9\alpha_{Y}^{3/2},\\&
C_{5}= 0.356,\\&
C_{6}=3.18373 \times10^{8} \alpha_{Y} , \\&
C_{7}=2.629\times 10^{22} \sqrt{\alpha_{Y}},\\& 
C_{8}=2.52\times 10^{27} \sqrt{\alpha_{Y}}.
\end{split}
\end{equation}

We choose the profile of temperature fluctuation, $\beta(x)$, to be Gaussian \cite{s3}
\begin{equation}
\beta(x)=\frac{\beta_{0}}{b\sqrt{2\pi}}\exp\left[-\frac{(x-x_{0})^{2}}{2b^{2}}\right], 
\end{equation}
where $\beta_{0}$ is the amplitude multiplying the normalized Gaussian distribution. We also choose the the profile of vorticity fluctuation to be the same as that of temperature fluctuation in order to produce maximal effect, that is \cite{s3} 
\begin{equation}
\omega(x)=\frac{\omega_{0}}{b\sqrt{2\pi}}\exp\left[-\frac{(x-x_{0})^{2}}{2b^{2}}\right],
\end{equation}
where $\omega_{0}=k^{\prime}v_{0}$, and $v_{0}$ is the amplitude of the initial velocity. In the next section we solve the set of coupled evolution equations (\ref{eq47}-\ref{eq49}), for various values of $\{\beta_0, v_0, x_0, b,k\}$, and use Eq.\ (\ref{mus}) to obtain all of the asymmetries and the hypermagnetic field amplitude in the temperature range $100\mbox{GeV} \le T\le 10\mbox{TeV}$. The initial condition for each of these quantities is zero. We also obtain and display the results for two sets of consecutive and opposite fluctuations.

\section{Numerical Solution}\label{x3}
In this section, we solve the set of coupled differential equations numerically, in the temperature range $100 \mbox{GeV}\leq T \leq 10\mbox{TeV}$. First, we solve the AMHD equations for the monochromatic Chern-Simons wave configuration of the hypermagnetic and velocity fields, and then we extend the analysis by considering the continuous spectra of the fields, as presented in Appendix~\ref{app-c}. We investigate the effects of overlapping transient fluctuations in the temperature of some matter degrees of freedom and vorticity of the plasma, on the generation and evolution of the hypermagnetic field and the matter-antimatter asymmetries \cite{s3}. We show that the vorticity and temperature fluctuations have the maximum effectiveness when they occur concurrently. Since our main purpose is to produce the hypermagnetic field and matter-antimatter asymmetries from zero initial values, we set $B_{Y}^{(0)}=0$, and  $\eta_{e_R}^{(0)}=\eta_{\mu_R}^{(0)}=\eta_{\tau_{R}}^{(0)}=0$.  

\begin{figure*}[]
\centering
\subfigure[]{\label{fig:figure:1w}
\includegraphics[width=.45\textwidth]{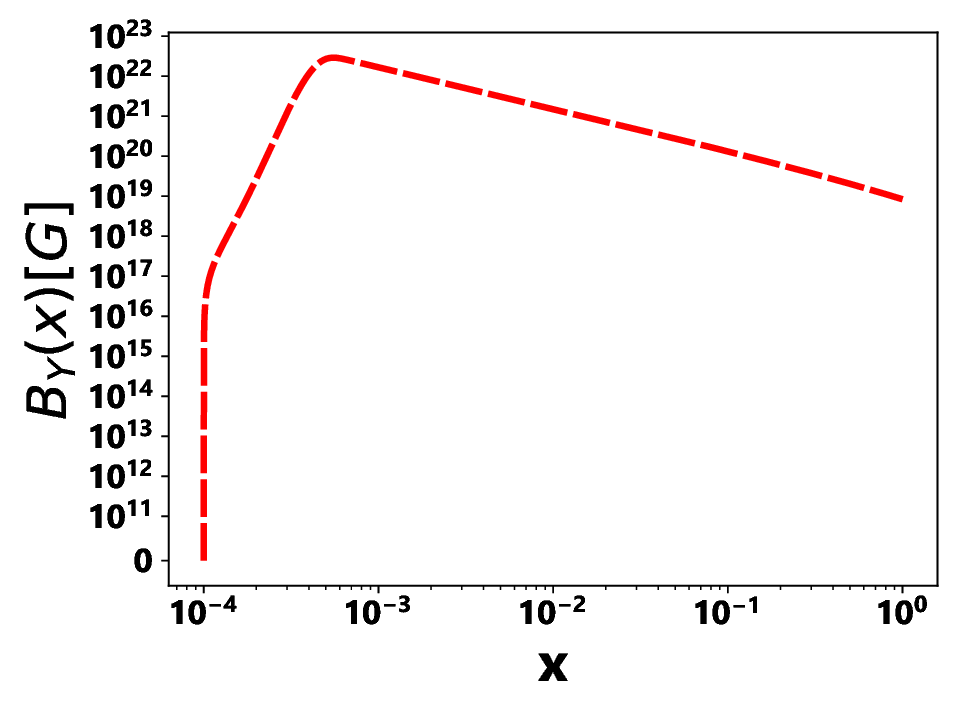}}
\hspace{8mm}
\subfigure[]{\label{fig:figure:2w} 
\includegraphics[width=.45\textwidth]{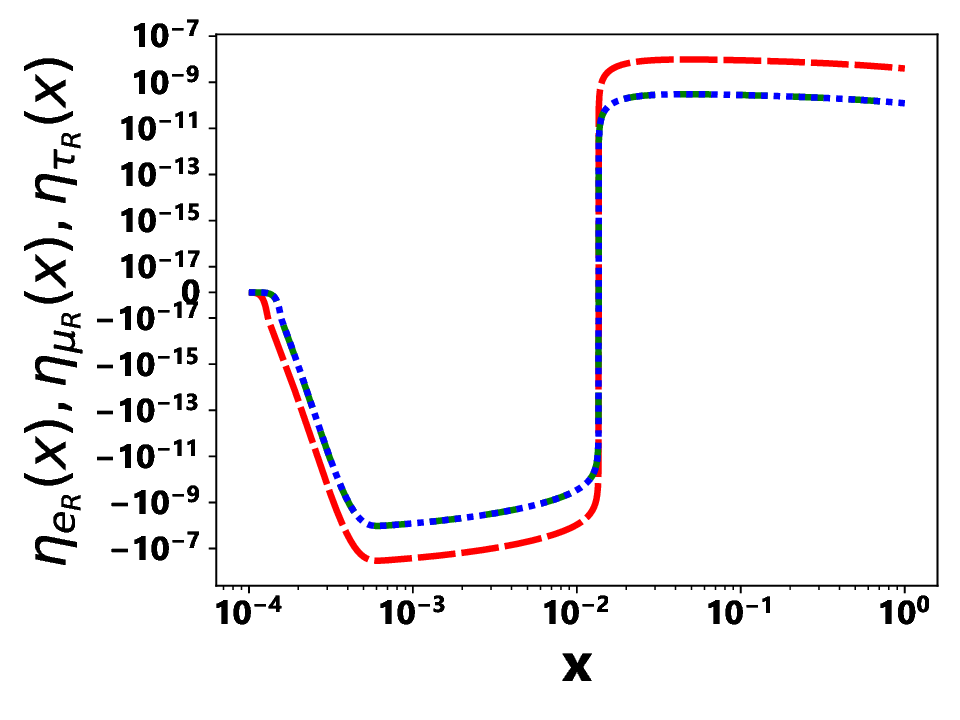}}
\hspace{8mm}
\subfigure[]{\label{fig:figure:3w} 
\includegraphics[width=.45\textwidth]{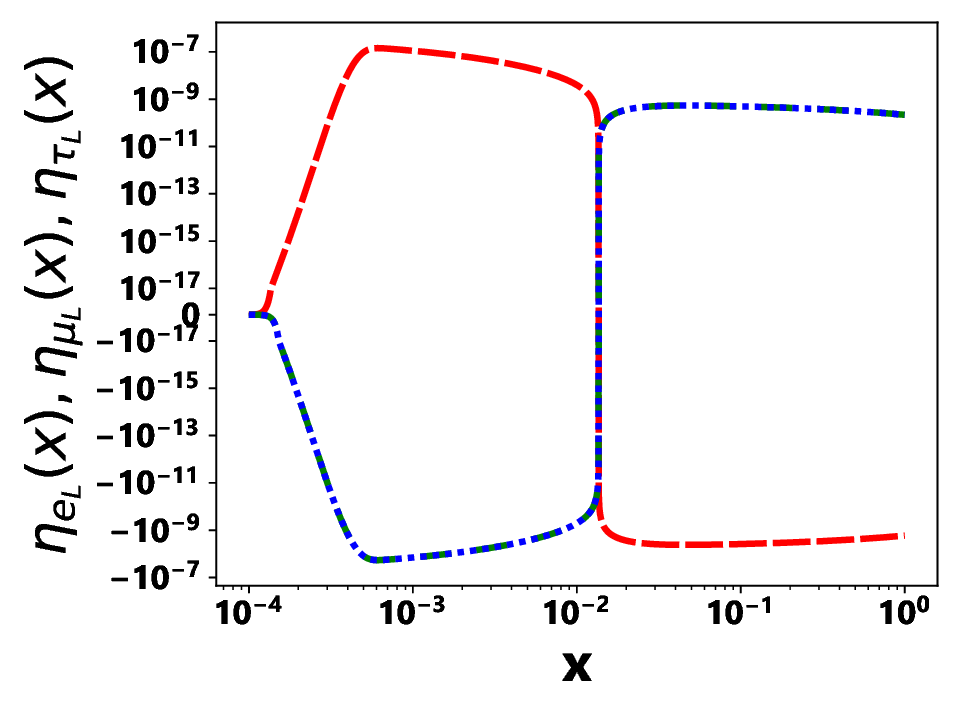}}
\hspace{8mm}
\subfigure[]{\label{fig:figure:4w} 
\includegraphics[width=.45\textwidth]{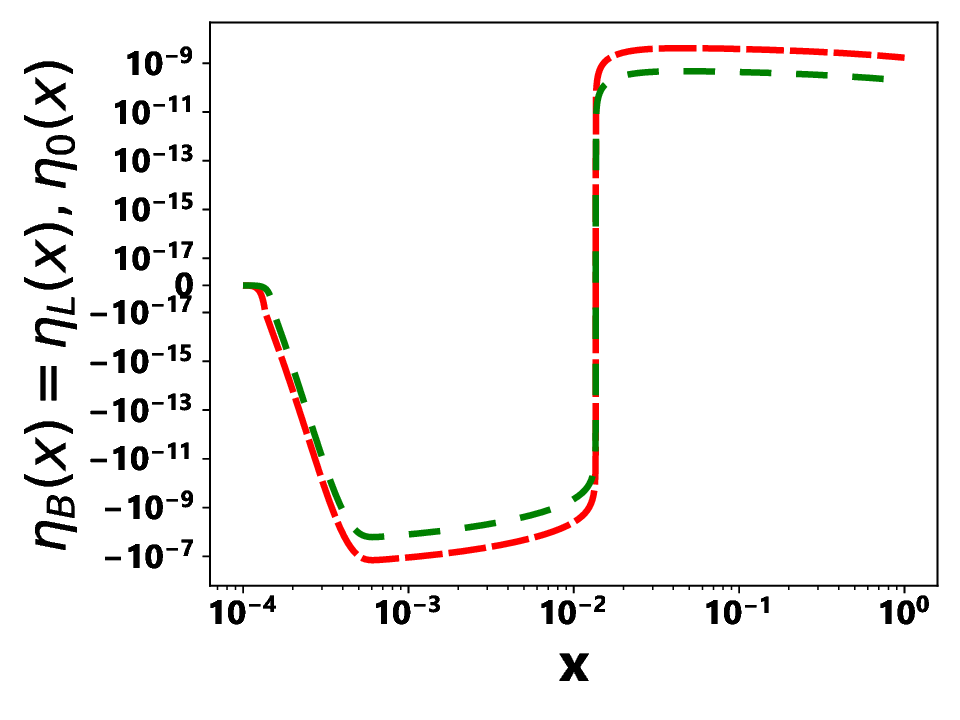}}
\hspace{8mm}
\subfigure[]{\label{fig:figure:5w} 
\includegraphics[width=.45\textwidth]{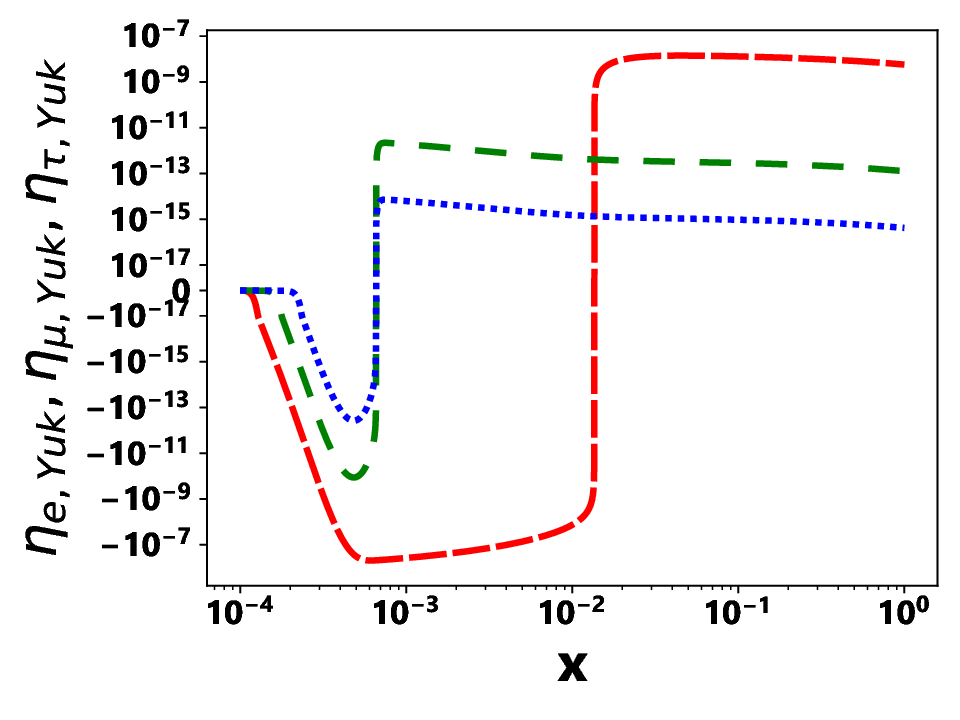}}
\caption{\footnotesize Time plots of (a) the hypermagnetic field amplitude $B_{Y}(x)$,  (b) the asymmetries of right-handed electron $\eta_{e_R}(x)$ (large-dashed-red),  right-handed muon $\eta_{\mu_R}(x)$ (dashed-green), and right-handed tau $\eta_{\tau_R}(x)$ (dotted-blue), (c) the asymmetries of left-handed electron $\eta_{e_L}$ (large-dashed-red), left-handed muon  $\eta_{\mu_L}(x)$ (dashed-green), and left-handed tau $\eta_{\tau_L}(x)$ (dotted-blue), (d) the baryon  and lepton asymmetry $\eta_B(x) = \eta_L(x)$ (large-dashed-red), and the Higgs asymmetry $\eta_{0}(x)$ (dashed-green), (e) the amounts of falling out of chemical equilibrium $\eta_{e,\mathrm{Yuk}}(x)$ (large-dashed-red), $\eta_{\mu, \mathrm{Yuk}}(x)$ (dashed-green), and $\eta_{\tau,\mathrm{Yuk}}(x)$ (dotted-blue), for the initial conditions $k=10^{-7}$, $B_{Y}^{(0)}=0$, $\eta_{e_R}^{(0)}=\eta_{\mu_R}^{(0)}=\eta_{\tau_R}^{(0)}=0$, $v_0=10^{-4}$, $\beta_{0}=5\times10^{-4}$, $x_{0}=4.5\times10^{-4}$, and $b=1\times10^{-4}$. } 
\label{fig1-w}
\end{figure*}

For our first case, we consider simultaneous fluctuations in the temperature of the right-handed electrons, and vorticity of the plasma \cite{s3}, and solve the set of evolution equations with the initial conditions, $v_0=10^{-4}$, $b=1\times10^{-4}$,  $x_{0}=4.5\times10^{-4}$, and $\beta_{0}=5\times10^{-4}$, and present the results in Fig.\ \ref{fig1-w}. Figure \ref{fig:figure:1w} shows that small overlapping temperature and vorticity fluctuations activate the CVE, leading to the generation of a strong helical hypermagnetic field, which grows to its maximum value of about $10^{22}$G and then begins decreasing due to the expansion of the Universe. We like to emphasize that the helicity of this hypermagnetic field is positive, contrary to our  previous work \cite{s3}. After the generation of the helical hypermagnetic field, the hypermagnetic helicity decays and produces the matter-antimatter asymmetries, all starting from zero initial values (see Figs.\ \ref{fig:figure:2w}, \ref{fig:figure:3w}, and \ref{fig:figure:4w}). Figs.\ \ref{fig:figure:2w} and \ref{fig:figure:3w} show that the asymmetry of any lepton other than the left-handed electron
has a peak with a negative value, and a positive final value\footnote{This is in contrast to our results in \cite{s3}, where $\eta_{e_L}$ initially became negative and finally attained a positive value, and $\eta_{e_R}$ was always positive.}. This is due to the fact that the hypermagnetic field makes the electron Yukawa process, which has the lowest rate, fall out of chemical equilibrium much more strongly than other processes. Therefore, the chiral electron asymmetries cannot be converted to each other effectively, and as a result, a large chiral asymmetry remains.

Figure ~\ref{fig:figure:4w} shows that the generated baryon asymmetry at the onset of the EWPT is acceptable, even though the weak sphaleron processes are in equilibrium\footnote{In our previous work \cite{s3}, $\eta_{B}$ was always positive, contrary to the results shown here.}. In fact, we have used this equilibrium condition as a constraint. This is equivalent to setting $(B+L)_L (t)= (B+L)_L (t_0)\equiv 0$, due to our initial conditions. Therefore, excess $(B+L)_R$ has been produced due to the anomalous processes.

Figure\ \ref{fig:figure:5w} shows $\eta_{e_{, \mathrm{Yuk}}}$, $\eta_{\mu_{,\mathrm{Yuk}}}$, and $\eta_{\tau_{, \mathrm{Yuk}}}$, which are measures for departure from equilibrium of Yukawa processes for electron, muon, and tau, respectively. The results show that the three lepton Yukawa interactions are initially in chemical equilibrium, however the generated strong hypermagnetic field forces them out of chemical equilibrium. Furthermore, the amount of falling out of equilibrium depends on the rate of the relevant Yukawa process such that the faster the process the less its departure from equilibrium. Since the electron Yukawa interaction has the smallest rate among all Yukawa processes, its departure from equilibrium is the largest.

\begin{figure*}[ht!]
\centering
\subfigure[]{\label{fig:figure:1wkr}
\includegraphics[width=.45\textwidth]{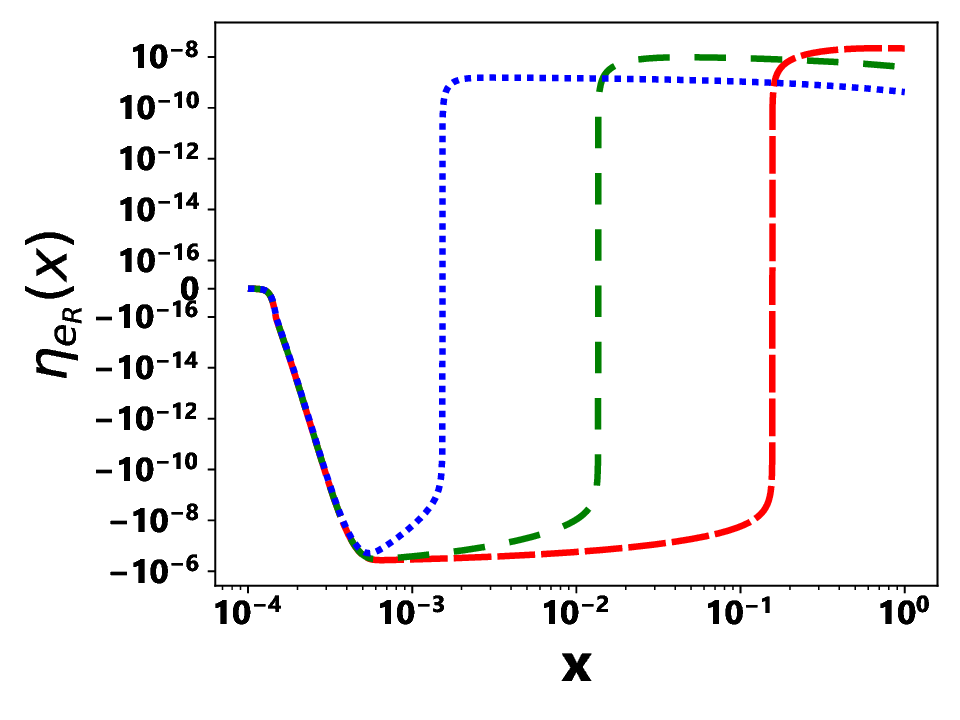}}
\hspace{8mm}
\subfigure[]{\label{fig:figure:2wkr} 
\includegraphics[width=.45\textwidth]{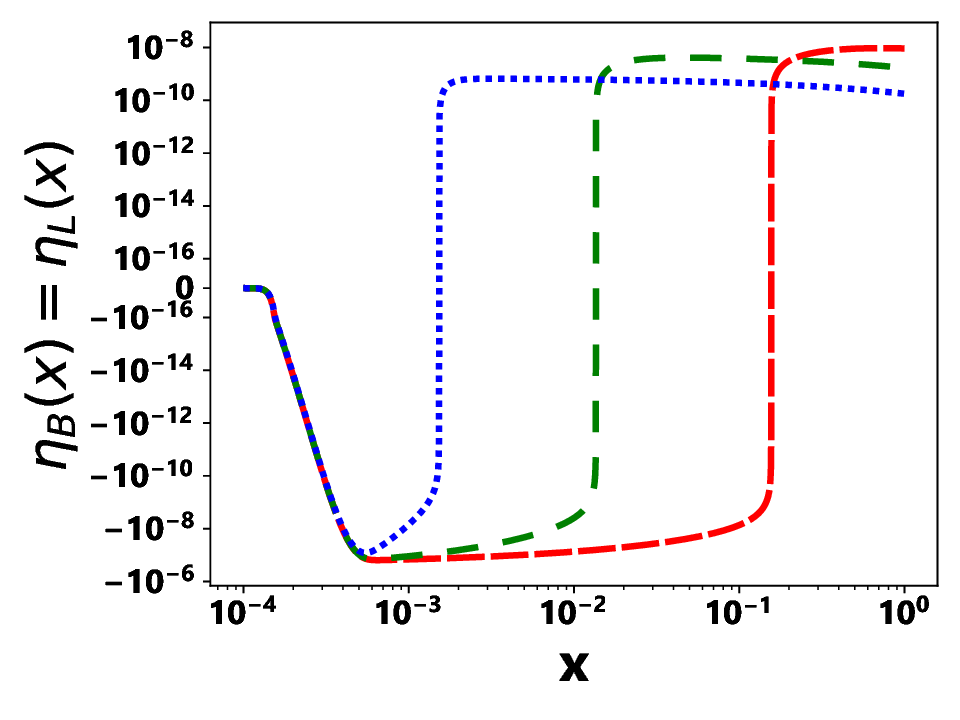}}
\caption{\footnotesize Time plots of (a) the right-handed electron asymmetry $\eta_{e_R}(x)$, and (b) the baryon  and lepton asymmetry $\eta_B (x)= \eta_L (x)$ for different values of $\lambda$. The initial condition are: $k=10^{-7}$, $B_{Y}^{(0)}=0$, $\eta_{e_R}^{(0)}=\eta_{\mu_R}^{(0)}=\eta_{\tau_R}^{(0)}=0$, $v_0=10^{-4}$,  $\beta_{0}=5\times10^{-4}$, $x_{0}=4.5\times10^{-4}$, and $b=1\times10^{-4}$. The dotted line (blue) is for $\lambda=10$, the dashed line (green) for $\lambda=1$, and the large-dashed line (red) for $\lambda=0.1$.}
\label{fig1-wkr}
\end{figure*}

As is well known, the ability of sphalerons to washout B+L is enhanced when all Yukawa processes are in equilibrium \cite{Turner}. Since, in the presence of the strong hypermagnetic field, the electron Yukawa process falls out of equilibrium more than other interactions, it plays an important role in the generation of matter-antimatter asymmetries.  To elucidate this statement, we investigate the effect of changing the rate of electron Yukawa interaction by considering  $\lambda\Gamma_{e}^0$ instead of $\Gamma_{e}^{0}$ in  Eq.\ (\ref{eq47}), where $\lambda\in \{0.1,1,10\}$. We solve the evolution equations with the  same initial conditions as before and present the results in Fig.\ \ref{fig1-wkr}. As can be seen, by decreasing the rate of electron Yukawa process $\lambda\Gamma_{e}^0$, the generated baryon and lepton asymmetries at the onset of the EWPT increase. 

\begin{figure*}[ht!]
%\centering
\subfigure[]{\label{fig:figure:221}
\includegraphics[width=.45\textwidth]{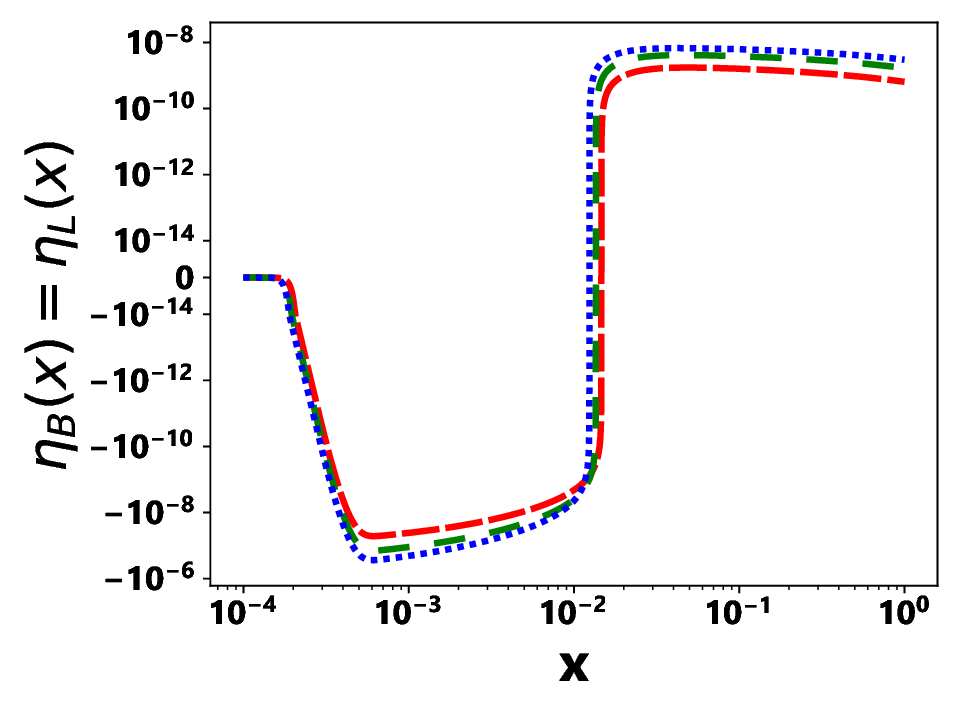}
}
\hspace{8mm}
\subfigure[]{\label{fig:figure:231}
\includegraphics[width=.45\textwidth]{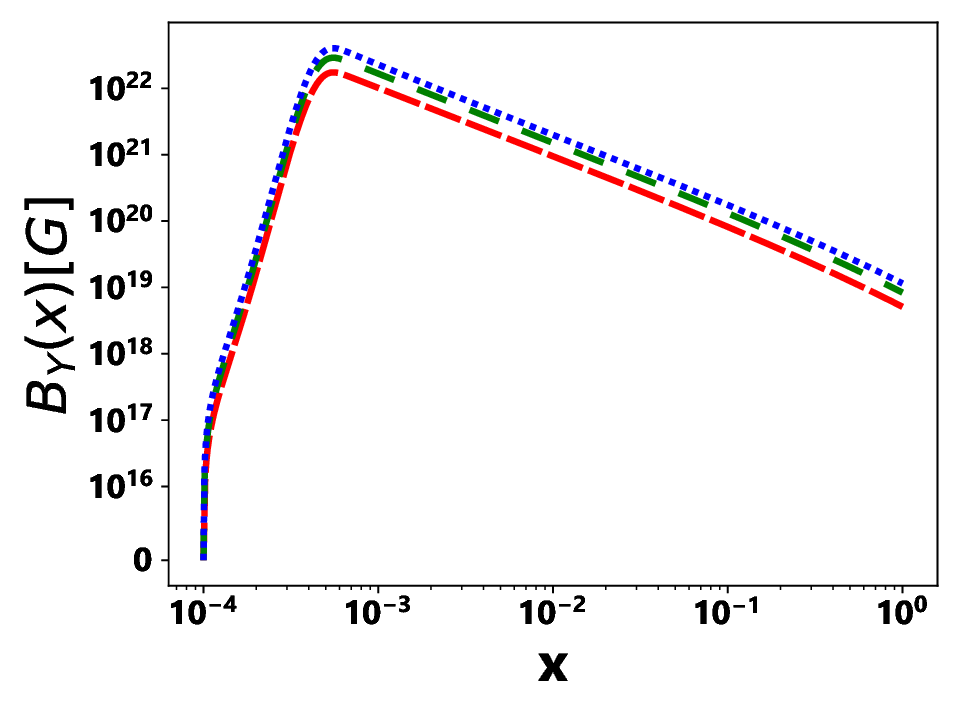}
}
\caption{\footnotesize Time plots of (a) the baryon and lepton asymmetry $\eta_{B}(x)=\eta_{L}(x)$, and (b) the hypermagnetic field amplitude $B_{Y}(x)$, for various values of the amplitude of temperature fluctuation of $e_{R}$. The initial conditions are: $k=10^{-7}$, $B_{Y}^{(0)}=0$, $\eta_{e_R}^{(0)}=\eta_{\mu_R}^{(0)}=\eta_{\tau_R}^{(0)}=0$, $v_0=10^{-4}$, $x_{0}=4.5\times10^{-4}$, and $b=1\times10^{-4}$. The large-dashed line (red) is for $\beta_{0}=3\times10^{-4}$, the dashed line (green) is for $\beta_{0}=5\times10^{-4}$, and the dotted line (blue) is for $\beta_{0}=7\times10^{-4}$.\\ }
\label{fig4}
\end{figure*}

For our second case, we investigate the effects of changing the amplitude of the temperature fluctuation and solve the coupled differential equations with the initial conditions, $B_{Y}^{(0)}=0$, $\eta_{e_R}^{(0)}=\eta_{\mu_R}^{(0)}=\eta_{\tau_R}^{(0)}=0$, $v_0=10^{-4}$, $b=1\times10^{-4}$, $x_{0}=4.5\times10^{-4}$, and three different values of $\beta_{0}=3\times10^{-4}, 5\times10^{-4}, \textrm{and } 7\times10^{-4}$, and present the results in Fig.\ \ref{fig4}. The results show that by increasing the amplitude of the temperature fluctuation, the maximum and the final values of the hypermagnetic field amplitude, and as a result, the matter-antimatter asymmetries increase, approximately as $\beta_{0}^2$. We have also investigated the effects of changing the amplitude of the vorticity fluctuation, and have found the similar results \cite{s3}.
%
%\textcolor{purple}{} \textcolor{green}{} \textcolor{green}{,} \textcolor{green}{, {\it i.e.},}
%\textcolor{red}{} \textcolor{orange}{()} \textcolor{magenta}{}
%

\begin{figure*}[ht!]
%\centering
\subfigure[]{\label{fig:figure:221w}
\includegraphics[width=.45\textwidth]{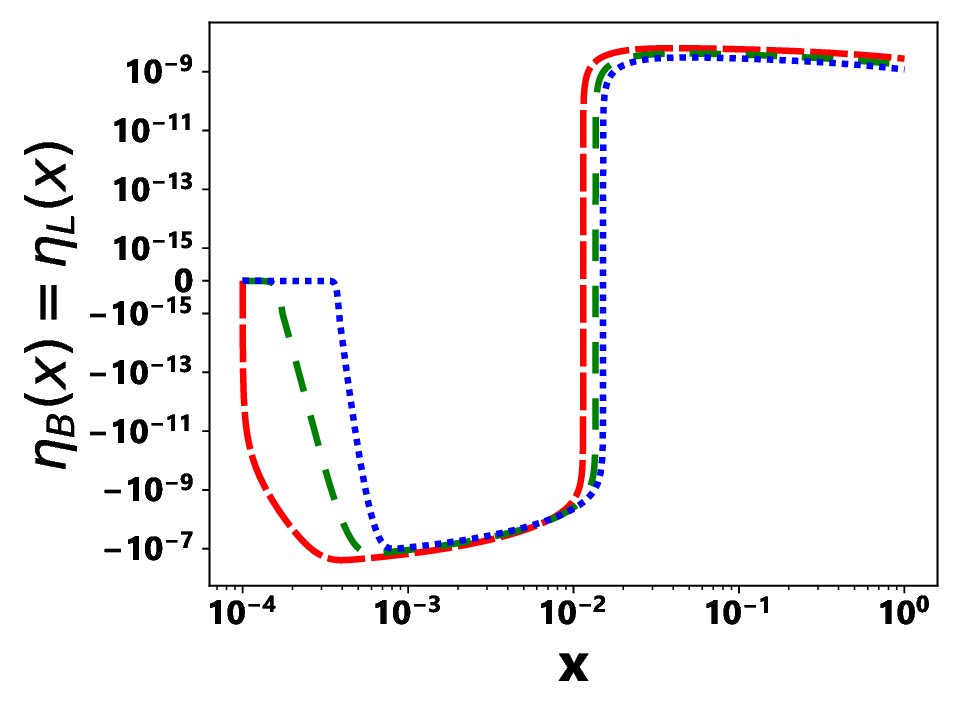}
}
\hspace{8mm}
\subfigure[]{\label{fig:figure:231w}
\includegraphics[width=.45\textwidth]{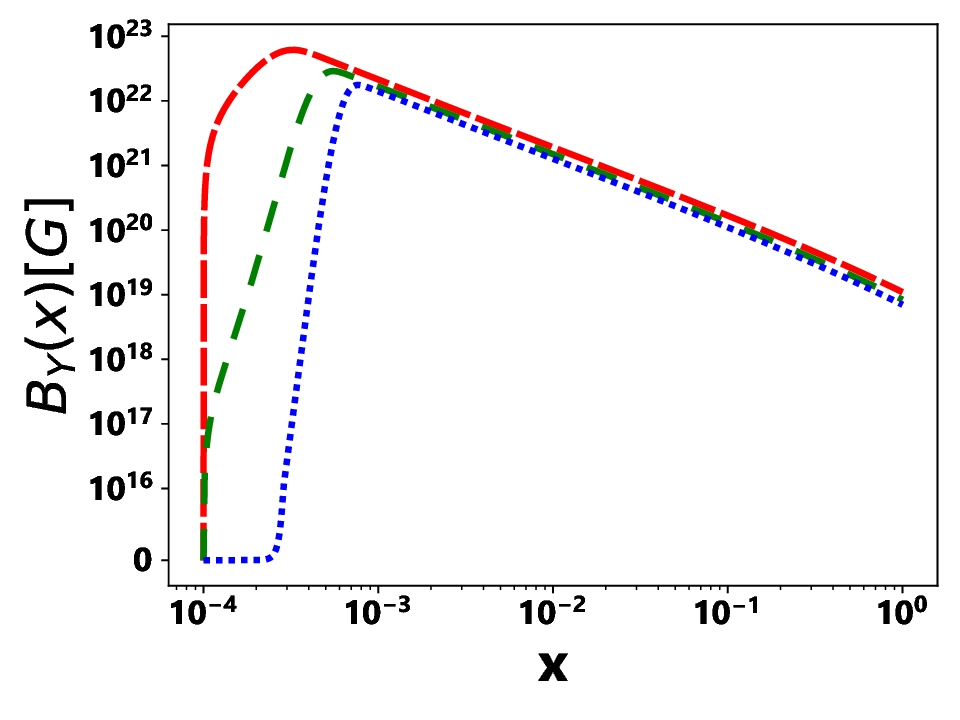}
}
\caption{\footnotesize Time plots of (a) the baryon and lepton asymmetry $\eta_{B}(x)=\eta_{L}(x)$, and (b) the hypermagnetic field amplitude $B_{Y}$, for various occurrence times of the fluctuations. The initial conditions are: $k=10^{-7}$, $B_{Y}^{(0)}=0$, $\eta_{e_R}^{(0)}=\eta_{\mu_R}^{(0)}=\eta_{\tau_R}^{(0)}=0$, $v_0=10^{-4}$, $\beta_{0}=5\times10^{-4}$, and $b=1\times10^{-4}$. The large-dashed line (red) is for $x_{0}=2.5\times10^{-4}$, the dashed line (green) is for $x_{0}=4.5\times10^{-4}$, and the dotted line (blue) is for $x_{0}=6.5\times10^{-4}$.\\ }
\label{fig5}
\end{figure*}

For our third case, we investigate the effects of changing the occurrence time of the  fluctuations and solve the coupled differential equations with the initial conditions, $B_{Y}^{(0)}=0$, $\eta_{e_R}^{(0)}=\eta_{\mu_R}^{(0)}=\eta_{\tau_R}^{(0)}=0$, $v_0=1\times10^{-4}$, $b=1\times10^{-4}$, $\beta_{0}=5\times10^{-4}$, and three different values of $x_{0}=2.5\times10^{-4}, 4.5\times10^{-4}, 6.5\times10^{-4}$, and present the results in Fig.\  \ref{fig5}. It can be seen that when the fluctuations occur at higher temperatures, the maximum and final values of the hypermagnetic field amplitude, and as a result, the matter-antimatter asymmetries increase. 

%\textcolor{magenta}{
%For our fourth case, we investigate the effects of changing the width of the fluctuation profiles and solve the coupled differential equations with the initial conditions, $B_{Y}^{(0)}=0$, $\eta_{e_R}^{(0)}=\eta_{\mu_R}^{(0)}=\eta_{\tau_R}^{(0)}=0$, $v_0=1\times10^{-4}$, $x_0=4.5\times10^{-4}$, $\beta_{0}=5\times10^{-4}$, and three different values of $b=2\times10^{-5}, 2\times10^{-4}, 2\times10^{-3}$, and present the results in Fig.\  \ref{fig1-wkq}. It can be seen that by decreasing the width of the Gaussian distribution, the maximum and final values of the hypermagnetic field amplitude, and as a result, the matter-antimatter asymmetries increase. }

%\textcolor{magenta}{} \textcolor{blue}{}

We have also investigated the case with different values for the width of the the Gaussian distribution $\beta[x(T)]$ and have found that by decreasing the width, the maximum and the final values of the hypermagnetic field amplitude, and as a result, the matter-antimatter asymmetries increase. Furthermore, we have examined two successive pulses with
opposite temperature and vorticity profiles to see by how much can the
second pulse negate the results of the first. To be specific, we assume
$\beta(x)=\beta_{+}(x)+\beta_{-}(x)$ and $v(x) = v_{+}(x) + v_{-}(x)$
where
$$\beta_{\pm}(x)=\pm\frac{\beta_{0}}{b\sqrt{2\pi}}  \exp[-\frac{(x-x_{0,\pm})^2}{2b^2}],$$
and
$$v_{\pm}(x)=\pm\frac{v_{0}}{b\sqrt{2\pi}}  \exp[-\frac{(x-x_{0,\pm})^2}{2b^2}].$$
We consider three cases in which the time separation of the
pulses $\Delta x_{0} =x_{0,+}-x_{0,-}$ are $5b$, $b$ and $0.1b$, where $ b=1\times10^{-4}$ denotes
the width of the pulses, and $x_{0,+}=4.5\times10^{-4}$ is the  occurrence time of first fluctuation. The final values of the asymmetries generated
are reduced, as compared to the single pulse case, by a factor
of about $5$, $50$ and $1000$, respectively. The final values of the
hypermagnetic field generated are reduced by the square root of
values stated above. It is interesting to note that even in the
case $\Delta x_{0} = 0.1b$, the model produced $ \eta_B\simeq 10^{-13}-10^{-14}$, a value
which can be increased easily by increasing $\beta_{0}$ and $v_0$ \cite{s3}.\\

%############################################################################################

\begin{figure*}[ht!]
	\centering
	\subfigure[]{\label{NewspecB1x}
		\includegraphics[width=.46\textwidth]{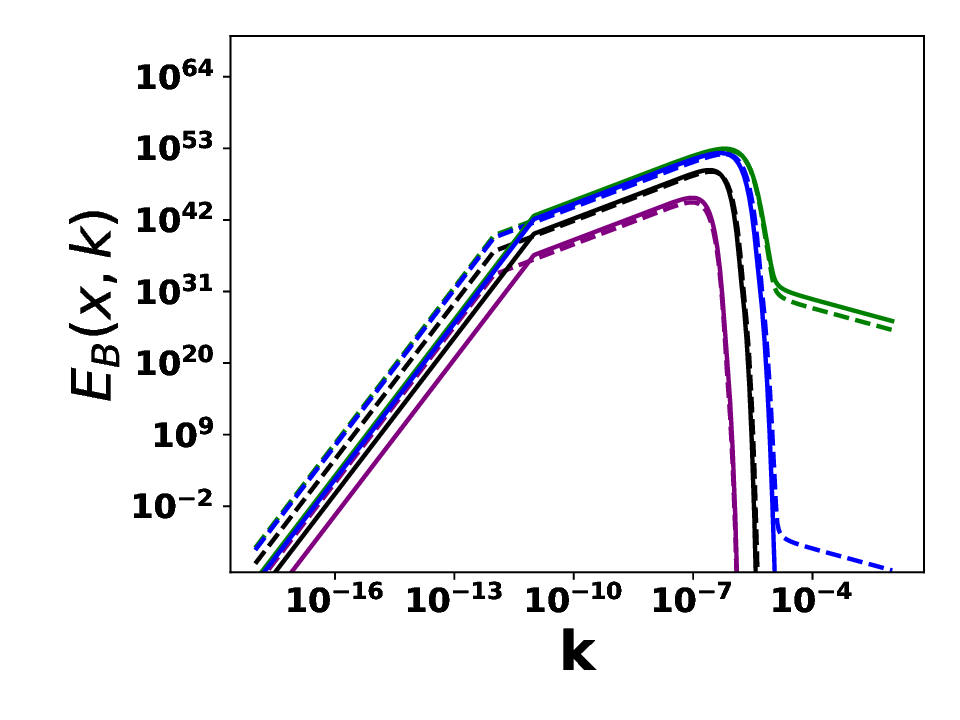}}
	\hspace{8mm}
	\subfigure[]{\label{NewspecH1x} 
		\includegraphics[width=.46\textwidth]{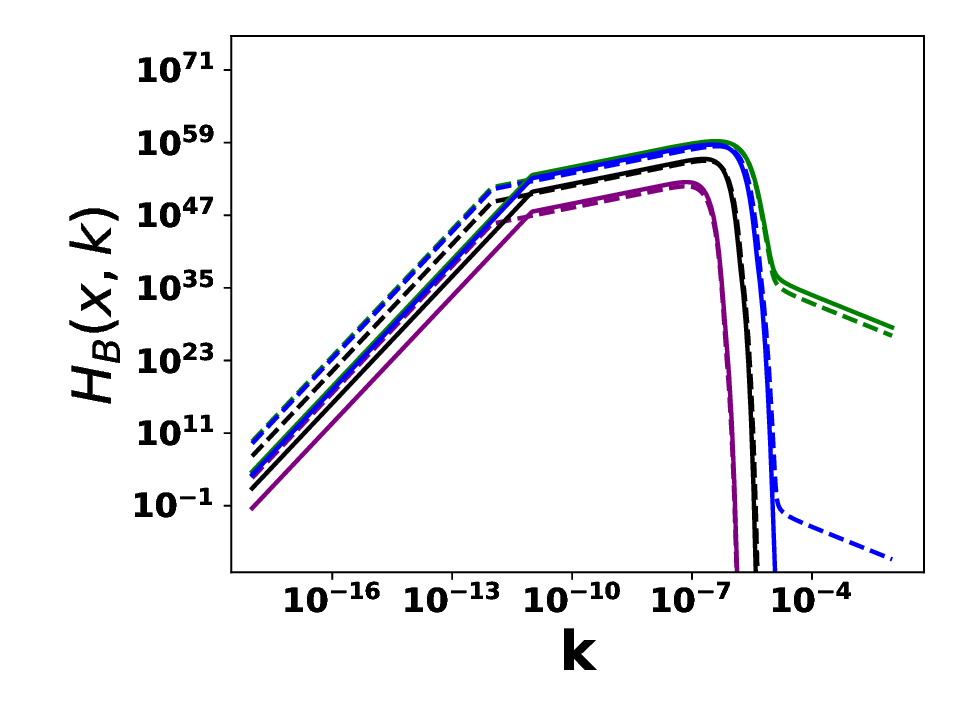}}
	\caption{\footnotesize Time evolution plots of (a) the hypermagnetic energy spectra $E_{B}(x,k)=k^2 M(x,k)/4\pi^2$ and (b) the hypermagnetic helicity spectra $H_{B}(x,k)=k M(x,k)/2\pi^2$ are shown for two values of the wave number $k_{\star}$. The initial conditions are as follows: $E_{B}(x_0,k) = 0$, $\eta_{e_R}^{(0)} = \eta_{\mu_R}^{(0)} = \eta_{\tau_R}^{(0)} = 0$, $v_0 = 10^{-3}$, $\beta_{0} = 5 \times 10^{-4}$, $x_{0} = 4.5 \times 10^{-4}$, and $b = 1 \times 10^{-4}$. The dashed lines represent $k_{\star} = 10^{-12}$, and the solid lines represent $k_{\star} = 10^{-11}$. The colors correspond to different values of $x$: green for $x = 9.6 \times 10^{-4}$, blue for $x = 1.4 \times 10^{-3}$, black for $x = 1.5 \times 10^{-2}$, and purple for $x = 1$, which is the final spectrum at $T=T_{\rm EW}$.}
	\label{fig6}
\end{figure*}

\begin{figure*}[ht!]
	\centering
	\subfigure[]{\label{Newbaryon1x}
		\includegraphics[width=.45\textwidth]{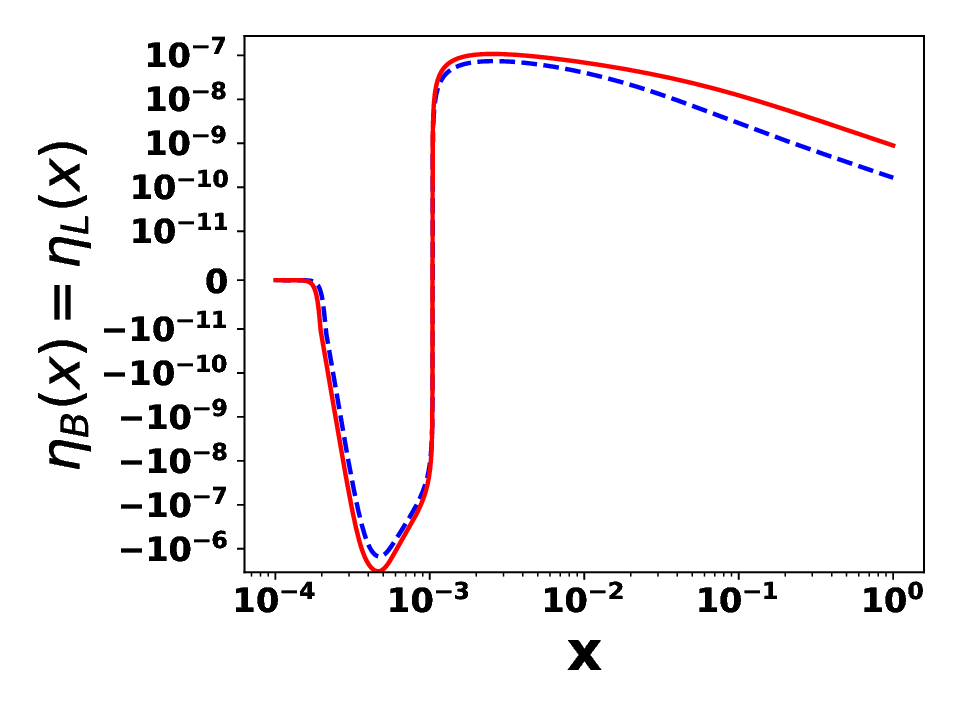}}
	\hspace{8mm}
	\subfigure[]{\label{NewBenergy1x} 
		\includegraphics[width=.45\textwidth]{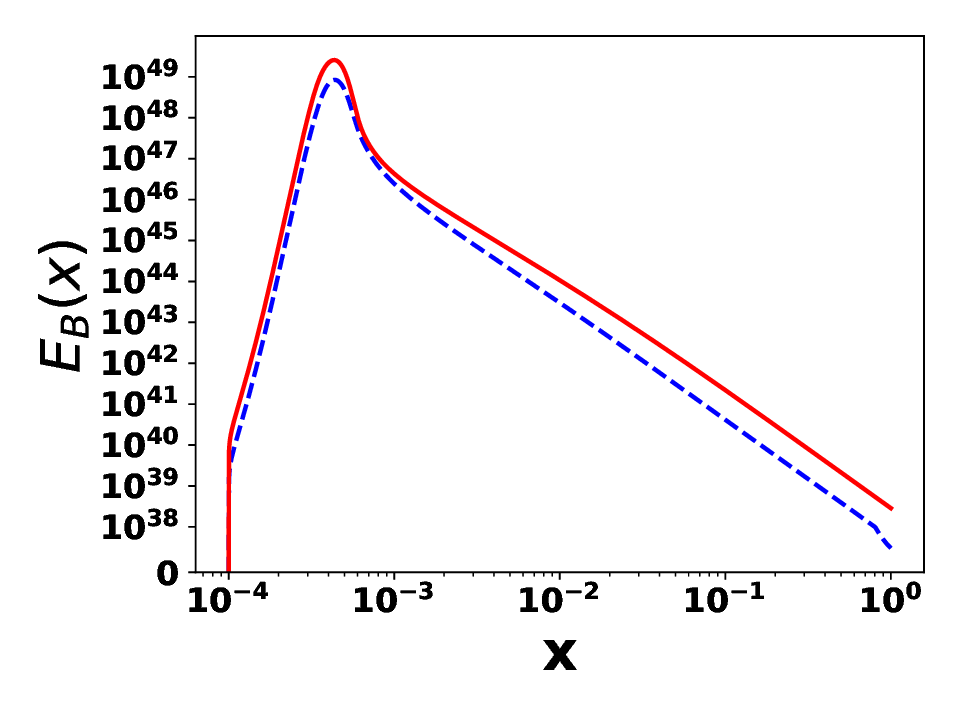}}
	\caption{\footnotesize Time evolution plots of (a) the baryon and lepton asymmetry $\eta_{B}(x)=\eta_{L}(x)$, and (b) the hypermagnetic field energy $E_{B}(x)=2B_Y^2(x)$, for two values of the $k_{\star}$. The initial conditions are: $E_{B}({x_0,k})=0$, $\eta_{e_R}^{(0)}=\eta_{\mu_R}^{(0)}=\eta_{\tau_R}^{(0)}=0$, $v_0=10^{-3}$, $\beta_{0}=5\times10^{-4}$,  $x_{0}=4.5\times^{-4}$, and $b=1\times10^{-4}$. The dashed line (blue) correspond to $k_{\star}=10^{-12}$ and the solid line (red) correspond to $k_{\star}=10^{-11}$.}
	\label{fig7}
\end{figure*}

For our fourth case, we investigate the simultaneous evolution of the hypermagnetic energy spectrum, $ E_B(x, k) $, and the hypermagnetic helicity spectrum, $ H_B(x, k) $, along with the generation of matter-antimatter asymmetries. To achieve this, we adopt a helical basis to extend the monochromatic Chern-Simons wave configurations of the hypermagnetic and velocity fields to continuous spectra (see Appendix~\ref{app-c}) \cite{s1}. We solve the coupled differential equations given in Eqs.~(\ref{eq49a}--\ref{eq49c}) with the following initial conditions: $ E_B(x_0, k) = 0 $, $ \eta_{e_R}^{(0)} = \eta_{\mu_R}^{(0)} = \eta_{\tau_R}^{(0)} = 0 $, $ v_0 = 1 \times 10^{-3} $, $ b = 1 \times 10^{-4} $, $ \beta_0 = 5 \times 10^{-4} $, and $ x_0 = 4.5 \times 10^{-4} $, for two different values of $ k_{\star} = 10^{-12}, 10^{-11} $ \cite{Dvornikov:2022gkf}. The transition wave number $k_{\star}$ denotes the scale separating the Batchelor and Kolmogorov regimes in the kinetic energy spectrum. The results are presented in Figs.~\ref{fig6} and ~\ref{fig7}. Note that both the hypermagnetic energy and helicity spectra are zero at the initial time, and they are generated from zero initial values by the chiral vortical effect (CVE) term originating from the fluctuations, which appears as the coefficient $C_7$ in the spectra evolution equation. Therefore, the initial spectrum for kinetic energy will generate the initial seed  for the hypermagnetic energy and helicity spectra.   
As can be seen from Fig.~\ref{fig6}, the log-log plot of the generated spectra for both fields show a sudden change of slope at $k=k_{\star}$ for all times, resulting from the change in the velocity spectrum at  $k_{\star}$. The amplitudes in both spectra decrease as a function of time due to the expansion and the magnetic diffusion. The latter is proportional to $k^4$ and, hence, much more effective for larger values of $k$.
Figure \ref{fig7} shows the time evolution of the baryon and total lepton asymmetry and the hypermagnetic energy $E_B(x)=B_Y^2(x)/2$. The results show that by increasing $k_{\star}$, the amplitudes for all these quantities increase, due to larger integrated kinetic energy, at all times including at the onset of the EWPT.

%%%%%%%%%%%%%%%%%%%%%%%%%%%%%%%%%%%%%%%%%%%%%%%%%%%%%%%%%%%%%%%%%%%
%%%%%%%%%%%%%%%%%%%%%%%%%%%%%%%%%%%%%%%%%%%%%%%%%%%%%%%%%%%%%%%%%%

Finally, we have checked the numerical accuracy of our program by plotting some of the quantities that we have set as constraints, {\it i.e.}, the hypercharge neutrality condition, the $B/3-L_i$ conditions, and the equilibrium condition for the weak sphaleron, and found that the absolute and relative uncertainties are of the order of $10^{-22}$ and $10^{-12}$, respectively.

\section{Conclusion}\label{x4}
In this work, we have presented a scenario for the generation of the hypermagnetic field and the matter-antimatter asymmetries all starting from zero initial values, in the presence of the weak sphaleron processes, and in the temperature range $100 \rm GeV\le T\le10\rm TeV$. We have considered all perturbative Yukawa interactions and nonperturbative Abelian and non-Abelian anomalous effects, as well as the contribution of all fermion and boson asymmetries in the anomalous transport effects, {\it i.e.}, the CVE and the CME. 

We have shown that overlapping of small transient fluctuations in the temperature of some matter degrees of freedom and vorticity of the plasma can activate the CVE producing a vortical current which can be the source for the hypermagnetic field, and as a result, matter-antimatter asymmetries \cite{s3}. Indeed, the baryon and lepton asymmetries and the hypermagnetic field helicity are produced, without B-L generation.
 We have shown that although $\rm (B+L)_L$ remains zero due to sphaleron processes being in equilibrium, excess $\rm (B+L)_R$ has been produced due to the anomalous processes.  Therefore, with our choice of initial conditions, acceptable baryon asymmetry can be generated at the onset of the EWPT. 
%
%\textcolor{purple}{} \textcolor{green}{} \textcolor{green}{,} \textcolor{green}{, {\it i.e.},}
%\textcolor{red}{} \textcolor{orange}{()} \textcolor{magenta}{}
%

Concurrently with the generation of the strong hypermagnetic field, the lepton Yukawa processes, which were initially in equilibrium, begin to fall out of equilibrium. The amount of falling out of chemical equilibrium depends on the rate of the interaction. That is, the smaller the rate, the larger the departure from equilibrium. Therefore, the Yukawa interaction of the electron falls out of chemical equilibrium more severely than other processes and hence it is an important process for generation of the $\rm B+L$ asymmetry. This is due to the fact that, in the absence of $\rm B-L$ asymmetry,  the weak sphaleron processes can wash out the asymmetries effectively when all Yukawa interactions are in equilibrium. 

We have also found that increasing the amplitude or decreasing the width of the Gaussian profile of the fluctuations, either in the temperature of the right-handed electron or in the vorticity of the plasma, leads to an increase in the maximum and final values of the hypermagnetic field amplitude and the matter-antimatter asymmetries.  We have also shown that the results for two sets of consecutive and opposite fluctuations only partially cancel each other, {\it i.e.}, baryon asymmetry and hypermagnetic field can still be generated. 
Furthermore, We have extended the monochromatic Chern-Simons wave configurations to continuous spectra of hypermagnetic and velocity fields. Comparing Fig.~\ref{fig7} with Fig.~\ref{fig4}, we note that the qualitative behavior of the generated baryon and lepton asymmetries, and hypermagnetic field energy in the continuous spectra case are similar to the corresponding results for the monochromatic Chern-Simons wave configuration case.

%Various observations indicate that the strength of the intergalactic magnetic fields is $B_0 \simeq 10^{-15}\text{G}$ \cite{Ando:2010rb, Essey:2010nd, Chen:2014rsa}, with scales as large as $\lambda_0 \simeq 1 \text{Mpc}$ and a non-vanishing helicity also inferred \cite{Chen:2014qva}. To compare the results that we have obtained for the hypermagnetic field at the onset of the EWPT to these observations, we could make the approximation that the hypermagnetic field is converted to magnetic field merely with a reduction factor $\cos \theta_{\rm W}$. Next, its evolution continues in the broken phase, in which usually the inverse cascade is invoked and helicity conservation is assumed \cite{n5}. The results are that, as an example, a magnetic field  with correlation length $\lambda(T_{\rm EW}) \simeq 2\pi/k T_{\rm EW}\simeq 4 \times 10^{-27} \text{ pc}$ and amplitude $ B(T_{\rm EW}) \simeq 10^{19} \text{G}$ evolves into one with $\lambda_0  \simeq 2 \times 10^{-4} \text{ pc}$ and $B{_0} \simeq 10^{-15} \text{ G}$ at the present. Evidently the amplitude is acceptable, but the scale is not. This generic problem for primordial magnetogenesis indicates that other mechanisms need to be considered to explain their large scale. More complex models, such as those incorporating turbulence-driven and anomaly-driven inverse cascade mechanisms, could be explored in both the symmetric and broken phases to enhance the scale of magnetic fields \cite{Boyarsky:2011uy, Brandenburg:2017rcb, Rogachevskii:2017uyc, Schober:2017cdw}.

Various observations indicate that the strength of the intergalactic magnetic fields is $B_0 \simeq 10^{-15}\text{G}$ \cite{Ando:2010rb, Essey:2010nd, Chen:2014rsa}, with scales as large as $\xi_B (T_{0}) \simeq 1 \text{Mpc}$ and a non-vanishing helicity also inferred \cite{Chen:2014qva}. The calculations of magnetic fields produced after inflation typically encounter the small-scale problem, where their comoving correlation lengths are significantly smaller than the scales of magnetic fields observed throughout the Universe \cite{n4,n5,aaaa,bbbb,cccc}. This small-scale problem is generally considered to be an open problem \cite{n4,n5,aaaa,bbbb,cccc}\footnote{In a recent study, it has been shown that if strong hypermagnetic fields are present at $T=10\rm TeV$, the results can be consistent with the Planck  observations \cite{Dvornikov:2022gkf}. }. The primordial magnetic fields are usually assumed to be statistically homogeneous and isotropic Gaussian random fields \cite{cccc}.
Assuming that the field spectrum is random, the correlation length of the magnetic field is inherently limited. The correlation length is then estimated by $\xi_B(T_{\rm EW}) \simeq 2\pi/k T_{\rm EW}\simeq 4 \times 10^{-23} \text{ pc}$ \cite{n5}.  However, in our first simple model, for example, the hypermagnetic field has a monopole Chern-Simons configuration , whose correlation function at EWPT and distance $r$ is $C(r)|_{T_{\rm EW}} = B(T_{\rm EW})^2 \cos(kr)$. Therefore, the correlation length for this simple configuration is limited only by the Hubble radius which is approximately $\xi_B (T_{\rm EW})\approx 3\text{cm}=9.7 \times10^{-19} \text{pc}$.
To compare the results that we have obtained for the hypermagnetic field at the onset of the EWPT to the observations, we make the approximation that the hypermagnetic field is converted to magnetic field merely with a reduction factor $\cos \theta_{\rm W}$. Next, its evolution continuous in the broken phase, in which usually the inverse cascade is invoked and helicity conservation is assumed \cite{n5}.  The results are that, as an example, a magnetic field with the above correlation length  and amplitude $ B(T_{\rm EW}) \simeq 10^{19} \text{G}$ that we have obtained here, evolves into one with $\xi_B (T_{0})   \simeq 4.85 \times 10^{4} \text{ pc}$ and $B{_0} \simeq 10^{-15} \text{ G}$ at the present. Evidently, the amplitude is acceptable, but the scale is two orders of magnitude smaller than the observed correlation length. In the broken phase, the Abelian anomaly does not violate baryon and lepton number conservation. Consequently, the existing matter-antimatter asymmetries remain unchanged during the inverse cascade that takes place following the electroweak phase transition. To increase the correlation length, more complex models, such as those incorporating turbulence-driven and anomaly-driven inverse cascade mechanisms, could be explored \cite{Boyarsky:2011uy, Brandenburg:2017rcb, Rogachevskii:2017uyc, Schober:2017cdw}.

Next we compare our results for matter-antimatter asymmetries with their corresponding observed values. The baryon-to-photon ratio is measured as $\eta=(n_{B}-\bar{n}_{B})/n_{\gamma}= (6.10 \pm 0.4) \times 10^{-10}$, where $n_{\gamma}$ is the photon number density \cite{Planck:2018vyg}. This asymmetry is equivalent to baryon-to-entropy ratio $\eta_{B}=(n_{B}-\bar{n}_{B})/s=(8.54\pm 0.056)\times 10^{-11}$. The results that we have calculated here depend on the parameters of our model, and the ones presented include $\eta_{B}\simeq 5\times10^{-10}$.

The lepton asymmetry of the Universe is significantly less constrained than the baryon asymmetry.
Recent determinations of the primordial abundance of ($^4$He) from the Big Bang Nucleosynthesis (BBN) indicate a favored positive asymmetry in electron neutrinos, expressed as $\eta_{\nu_e} (T_{\rm BBN}) \approx 1.75 \times 10^{-3}$ \cite{Matsumoto:2022tlr, Kawasaki:2022hvx,Gelmini:2020ekg}. Since charge neutrality requires that the asymmetry of the charged leptons be equal to the baryon asymmetry, it is assumed that the three active neutrinos carry the large lepton asymmetry.
Assuming equality between asymmetries of three neutrino flavors, due to neutrino oscillations, we can estimate the lepton asymmetry at the BBN epoch $\eta_L(T_{\rm BBN}) \approx 3 \eta_{\nu_e} (T_{\rm BBN})\approx 5.3 \times 10^{-3}$. We can convert our results for the lepton asymmetry at $T_{\rm EW}$ to $T_{\rm BBN}$ as,
$\eta_{L}^{\rm Our}(T_{\rm BBN})=5\times10^{-10}\frac{g^*(T_{\rm EW})}{g^*(T_{\rm BBN})}=4.965\times 10^{-9}$.
By adjusting the parameters of our model, the generated baryon asymmetry can be consistent with the existing constraint, while the generated lepton asymmetry is always equal to the baryon asymmetry. Therefore, the observed excess in lepton asymmetry necessitates an additional mechanism for its generation.

\appendix
\section{Anomalous versus conserved currents before the EWPT}\label{xx}
In the expanding Universe and before the EWPT, chiral fermionic currents have the following anomalies \footnote{As usual, the perturbative tree level chirality-flip processes are dropped.} \cite{a1,a2,a3,h1}:
\begin{equation}\label{er}
\nabla_{\mu} j_{{e}_{R}^i}^{\mu}=-\frac{1}{4}(Y_{R}^{2})\frac{g'^{2}}{16 \pi^2}Y_{\mu\nu}\tilde{Y}^{\mu\nu},
\end{equation}
\begin{equation}
\nabla_{\mu} j_{{l}_{L}^i}^{\mu}=\frac{1}{4}(N_{w}Y_{L}^{2})\frac{g'^{2}}{16\pi^2}Y_{\mu\nu}\tilde{Y}^{\mu\nu}+\frac{1}{2}\frac{g^{2}}{16\pi^2}W_{\mu\nu}^{a}\tilde{W}^{a\, \mu\nu},
\end{equation}
\begin{equation}
\nabla_{\mu} j_{{d}_{R}^{i}}^{\mu}=-\frac{1}{4}(N_{c}Y_{Q}^{2})\frac{g'^{2}}{16\pi^2}Y_{\mu\nu}\tilde{Y}^{\mu\nu}-\frac{1}{2}\frac{g_{s}^{2}}{16\pi^2}G_{\mu\nu}^{A}\tilde{G}^{A\, \mu\nu},
\end{equation}
\begin{equation}
\nabla_{\mu} j_{{u}_{R}^{i}}^{\mu}=-\frac{1}{4}(N_{c}Y_{Q}^{2})\frac{g'^{2}}{16\pi^2}Y_{\mu\nu}\tilde{Y}^{\mu\nu}-\frac{1}{2}\frac{g_{s}^{2}}{16\pi^2}G_{\mu\nu}^{A}\tilde{G}^{A\, \mu\nu},
\end{equation}
\begin{equation}
\nabla_{\mu}j_{q_L^i}^{\mu}=\frac{1}{4}(N_{c}N_{w}Y_{Q}^{2})\frac{g'^{2}}{16\pi^2}Y_{\mu\nu}\tilde{Y}^{\mu\nu}+\frac{1}{2}(N_{c})\frac{g^{2}}{16\pi^2}W_{\mu\nu}^{a}\tilde{W}^{a\, \mu\nu}+\frac{1}{2}(N_{w})\frac{g_{s}^{2}}{16\pi^2}G_{\mu\nu}^{A}\tilde{G}^{A\, \mu\nu},
\end{equation} 
%
%\textcolor{purple}{} \textcolor{green}{} \textcolor{green}{,} \textcolor{green}{, {\it i.e.},}
%\textcolor{red}{} \textcolor{orange}{()} \textcolor{magenta}{}
%
where $\nabla_{\mu}$ is the covariant derivative with respect to the FRW metric, `$i$' is the generation index, $ j_{{e}_{R}^i}^{\mu}$ ($ j_{{l}_{L}^i}^{\mu}=j_{{e}_{L}^i}^{\mu}+j_{{\nu}_{L}^i}^{\mu}$) is the right-handed singlet (left-handed doublet) lepton current, $ j_{{d}_{R}^i}^{\mu}$ ($ j_{{u}_{R}^i}^{\mu}$) is the right-handed down (up) singlet quark current, and $ j_{q_L^i}^{\mu}=j_{u_L^i}+j_{d_L^i}$ is the left-handed doublet quark current. Furthermore, $\rm N_{c}=3$ and $\rm N_{w}=2$ are the corresponding ranks of the non-Abelian $\rm SU(3)$ and $\rm SU_{L}(2)$ gauge groups, and $G_{\mu \nu}^{A}$, $W_{\mu \nu}^{a}$, and $Y_{\mu \nu}$ are the field strength tensors of the ${\rm SU}(3)$, ${\rm SU}_\textrm{L}(2)$, and ${\rm U}_\textrm{Y}(1)$ gauge groups with the fine structure constants $\frac{g_{s}^2}{4\pi}$, $\frac{g^2}{4\pi}$, and $\frac{g'^2}{4\pi}$, respectively. Moreover, the relevant hypercharges are  
\begin{equation}\label{eqds1}
\begin{split}
&Y_{L}=-1, \quad Y_{R}=-2, \quad Y_{Q}=\frac{1}{3}, \quad Y_{u_{R}}=\frac{4}{3}, \quad Y_{d_{R}}=-\frac{2}{3}.
\end{split}	
\end{equation} 
These anomaly equations show the following: The Abelian anomaly violates the conservation of all chiral lepton and quark currents, due to the chiral coupling of the $\rm U_{Y}(1)$ gauge fields to the fermions. The non-Abelian $\rm SU_{L}(2)$ gauge fields only couple to the left-handed fermions, so they violate the conservation of the left-handed lepton and quark currents. 
The non-Abelian $\rm SU(3)$ gauge fields only couple to the quarks and its sphaleron only changes their chiralities but respects the baryon current conservation \cite{h1,qn1}. 
The divergences of the total baryon and lepton currents, obtained from the anomaly equations, are as follows \cite{h1}:
\begin{equation}\label{eq.as1}
\begin{split}
&\nabla_{\mu}j^{\mu}_{\rm B}= \nabla_{\mu}\left[\frac{1}{N_c} \sum_{i} \left( j_{q_L^i}^{\mu} + j^{\mu}_{u_R^i} + j^{\mu}_{d_R^i} \right)\right]=n_G\left( \frac{g^2}{16 \pi^2} W_{\mu \nu}^a \tilde{W}^{a \, \mu \nu} - \frac{g^{\prime \, 2}}{16 \pi^2} Y_{\mu \nu} \tilde{Y}^{\mu \nu} \right),\\&
\nabla_{\mu}j^{\mu}_{\rm L} = \nabla_{\mu}\left[\sum_{i} \left( j^{\mu}_{l_L^i} + j^{\mu}_{e_R^i}  \right)\right]=n_G\left( \frac{g^2}{16 \pi^2} W_{\mu \nu}^a \tilde{W}^{a \, \mu \nu} - \frac{g^{\prime \, 2}}{16 \pi^2} Y_{\mu \nu} \tilde{Y}^{\mu \nu} \right), 
\end{split}
\end{equation}
where $n_G$ is the number of generations. Although, both of the baryonic and leptonic currents are separately anomalous,  Equation (\ref{eq.as1}) shows that their difference is anomaly free, $\nabla_{\mu}(j^{\mu}_{B-L})=0$, i.e., $B -L=$ constant.
In contrast to  $j^{\mu}_{B-L}$, the sum of the baryonic and leptonic currents $j^{\mu}_{B+L}\equiv j^{\mu}_{B}+j^{\mu}_{L}$ is anomalous,
\begin{equation}\label{eq.as2}
\begin{split}
&\nabla_{\mu}j^{\mu}_{\rm B+L}= 2n_G\left( \frac{g^2}{16 \pi^2} W_{\mu \nu}^a \tilde{W}^{a \, \mu \nu} - \frac{g^{\prime \, 2}}{16 \pi^2} Y_{\mu \nu} \tilde{Y}^{\mu \nu} \right).
\end{split}
\end{equation}
Eq.\ (\ref{eq.as2}) shows the non-conservation of $\rm B+L$ via the non-Abelian $\textrm{ SU}_\textrm{L}(2)$, and the Abelian $\textrm{U}_\textrm{Y}(1)$ gauge fields.

The total hypercharge current, including the contributions of Higgs and all chiral fermions, can be obtained as \cite{h1}
\begin{equation}\label{eq.jy}
J_{Y}^{\mu} = \sum_{i=1} ^{n_G}\left[ Y_Q j_{q_L^i}^{\mu} + Y_{u_R} j_{u_R^i}^{\mu} + Y_{d_R} j_{d_R^i}^{\mu} + Y_{L} j_{l_L^i}^{\mu} + Y_{R} j_{e_R^i}^{\mu}+ Y_{\varPhi}j_{\varPhi}^{\mu}\right],  
\end{equation}
where $j_{\varPhi}^{\mu}=j_{\varPhi^+}^{\mu}+j_{\varPhi^0}^{\mu}$ and $Y_{\varPhi}=1$ are the doublet current and the hypercharge of the Higgs boson, respectively. One can explicitly show that the total hypercharge current is divergence free, $\nabla_{\mu}J_{Y}^{\mu}=0$, as expected. In Sec.\ \ref{x2} we will use the anomaly free currents $j^{\mu}_{B-L}$ and $J_{Y}^{\mu}$ to obtain the corresponding conservation laws.

\section{Anomalous magnetohydrodynamics }\label{x1b}
Taking the CME and the CVE into account, the magnetohydrodynamics (MHD) is generalized to the anomalous magnetohydrodynamics (AMHD). The corresponding equations for a hypercharge neutral plasma in the expanding Universe are given as \cite{Giovannini-2013oga,Giovannini-1997eg,Giovannini-2015aea,Giovannini2000}
\begin{equation}\label{eq1}
\frac{1}{R}\vec{\nabla} .\vec{E}_{Y}=0,\qquad\qquad\frac{1}{R}\vec{\nabla}.\vec{B}_{Y}=0,
\end{equation}
\begin{equation}\label{eq2}
\frac{1}{R}\vec{\nabla}\times\vec{ E}_{Y}+\left(\frac{\partial \vec{B}_{Y}}{\partial t}+2H\vec{B}_{Y}\right)=0,
\end{equation}
\begin{equation}\label{eq3}
\begin{split}
\frac{1}{R}\vec{\nabla}\times\vec{B}_{Y}-&\left(\frac{\partial \vec{E}_{Y}}{\partial t}+2H\vec{E}_{Y}\right)=\vec{J}\\&=\vec{J}_{\mathrm{Ohm}}+\vec{J}_{\mathrm{cv}}+\vec{J}_{\mathrm{cm}},
\end{split}
\end{equation}
\begin{equation}\label{eq2.3}
\vec{J}_{\mathrm{Ohm}}=\sigma\left(\vec{E}_{Y}+\vec{v}\times\vec{B}_{Y}\right),
\end{equation}
\begin{equation}\label{eq3.1}
\vec{J}_{\mathrm{cv}}=c_{\mathrm{v}}\vec{\omega},
\end{equation}
\begin{equation}\label{eq3.2}
\vec{J}_{\mathrm{cm}}=c_{\mathrm{B}}\vec{B}_{Y},
\end{equation}
where $\vec{\nabla}$ is the covariant derivative with respect to the FRW metric, $R$ is the scale factor, $H=\dot{R}/R$ is the Hubble parameter, $\sigma$ is the electrical hyperconductivity of the plasma, $\vec{v}$  is the bulk velocity, and $\vec{\omega}=\frac{1}{R}\vec{\nabla}\times\vec{v}$ is the vorticity of the plasma. 
Furthermore, the hypercharge chiral magnetic coefficient $c_{\mathrm{B}}$ and the chiral vortical coefficient $c_{\mathrm{v}}$, in the symmetric phase, are given as \cite{s1}

\begin{equation}\label{eq26} 
\begin{split} 
&c_{\mathrm{B}}(t)=\frac{-g'^{2}}{8\pi^{2}} \sum_{i=1}^{n_{G}}\left[-2\mu_{e_R^i}+\mu_{e_L^i}-\frac{2}{3}\mu_{d_R^i}-\frac{8}{3}\mu_{u_R^i}+\frac{1}{3}\mu_{Q^{i}}\right], 
\end{split}
\end{equation}
\begin{equation}\label{eq24}
\begin{split}
c_{\mathrm{v}}(t)=&\sum_{i=1}^{n_{G}}\Big[\frac{g'}{24}\left(T_{e_R^i}^2-T_{e_L^i}^2+T_{d_R^i}^{2}-2T_{u_R^i}^{2}+T_{Q^{i}}^{2}\right)\\&+\frac{{g'}}{8\pi^{2}}\left(\mu_{e_R^i}^2-\mu_{e_L^i}^2+\mu_{d_R^i}^{2}-2\mu_{u_R^i}^{2}+\mu_{Q^i}^{2}\right)\Big].
\end{split} 
\end{equation}

Using the relations expressed in Eq.\ (\ref{mus}), $c_{\mathrm{B}}$ simplifies to
\begin{equation}\label{eq27}
\begin{split}
c_{\mathrm{B}}(t)=\frac{g'^{2}}{4\pi^{2}} \left(c_{e, \mathrm{Yuk}}+c_{\mu, \mathrm{Yuk}}+c_{\tau, \mathrm{Yuk}}\right)=\frac{g'^{2}}{4\pi^{2}}c_{T},
\end{split}	
\end{equation}
where $c_{e, \mathrm{Yuk}}\equiv\mu_{e_{R}}-\mu_{e_{L}}+\mu_{0}$, $c_{\mu, \mathrm{Yuk}}\equiv\mu_{\mu_{R}}-\mu_{\mu_{L}}+\mu_{0}$, and  $c_{\tau, \mathrm{Yuk}}\equiv\mu_{\tau_{R}}-\mu_{\tau_{L}}+\mu_{0}$ are measures for departure from chemical equilibrium for the Yukawa interactions of the electron, muon, and tau, respectively \cite{sh3}. Moreover, using Eq.\ (\ref{mus}) and assuming that the temperature fluctuation occurs only for the right-handed electron, $c_\mathrm{v}$ simplifies to  \cite{s3}
\begin{equation}\label{eq28}
c_{\mathrm{v}}(t)=\frac{g'}{24}\left(\Delta T^{2}\right)+\frac{g'}{2\pi^{2}}\left(\mu_{e}\mu_{e}^{5}+\mu_{\mu}\mu_{\mu}^{5}+\mu_{\tau}\mu_{\tau}^{5}-\frac{3}{8}\mu_{B}\mu_{0}-\frac{3}{4}\mu_{0}^2\right),
\end{equation}
where $\mu_{e}$, $\mu_{\mu}$, $\mu_{\tau}$ ($\mu_{e}^{5}$, $\mu_{\mu}^{5}$, $\mu_{\tau}^{5}$) are vector (axial-vector) chemical potential of the electron, muon, and tau, respectively.\footnote{The vector chemical potential is $\mu=(\mu_{R}+\mu_{L})/2$ and the axial-vector chemical potential is $\mu^{5}=(\mu_{R}-\mu_{L})/2$.}  %$\Delta\mu^{2}=\left(\mu_{e}\mu_{e}^{5}+\mu_{\mu}\mu_{\mu}^{5}+\mu_{\tau}\mu_{\tau}^{5}-\frac{3}{8}\mu_{B}\mu_{0}-\frac{3}{4}\mu_{0}^2\right)$,  %$\Delta\mu^{2}=\left(\mu_{e_{R}}^{2}-\mu_{e_{L}}^{2}+\mu_{\mu_{R}}^{2}-\mu_{\mu_{L}}^{2}+\mu_{\tau_{R}}^{2}-\mu_{\tau_{L}}^{2}-\frac{3}{2}\mu_{B}\mu_{0}-3\mu_{0}^2\right)\\=\left(\eta_{e}\eta_{e}^{5}+\eta_{\mu}\eta_{\mu}^{5}+\eta_{\tau}\eta_{\tau}^{5}-aaa\eta_{B}\right) $,
Furthermore, $\Delta T^{2}=T_{e_{R}}^{2}-T_{e_{L}}^{2}=T^{2} \beta[x (T)]$ is the temperature fluctuation, $\beta [x(T)]$ is an arbitrary profile function of temperature which will be specified later, $x(T)=t(T)/t_\mathrm{EW}=\left(T_\mathrm{EW}/T\right)^{2}$ is given by the Friedmann law, $t_\mathrm{EW}=M_{0}/2T_\mathrm{EW}^{2}$, $M_{0}=M_\mathrm{Pl}/1.66\sqrt{g^{*}}$ is the reduced Planck mass, and $T_{e_{L}}=T$ is the equilibrium temperature of the thermal bath \cite{s3}.
%
%\textcolor{purple}{} \textcolor{green}{} \textcolor{green}{,} \textcolor{green}{, {\it i.e.},}
%\textcolor{red}{} \textcolor{orange}{()} \textcolor{magenta}{}
%

Since the vorticity depends on the curl of velocity as $\vec{\omega}=\frac{1}{R}\vec{\nabla} \times \vec{v}$, in analogy to the hypermagnetic field $\vec{B}_Y=\frac{1}{R}\vec{\nabla} \times \vec{A}_Y$, it is also divergence free, {\it i.e.}, $\vec{\nabla}.\vec{\omega}=0$. Moreover, incompressibility assumption of the plasma leads to the condition $\vec{\nabla}.\vec{v}=0$ \cite{pav1}.\footnote{Here, for brevity, we do not present the energy and momentum conservation equations and only use the result obtained in Ref.\ \cite{s3}.} In the following, we choose the fully helical Chern-Simons configurations for the vector potentials of the hypermagnetic and vorticity fields as  \cite{ruba1,Giovannini2000,Joyce-1997uy,Giovannini-2015aea,Giovannini-1997eg,Giovannini-2013oga}
\begin{equation}\label{eq10}
\begin{split}
\vec{A}_{Y}=A(t)\left(\sin kz , \cos kz, 0\right),
\end{split}
\end{equation}
and
\begin{equation}\label{eq11}
\begin{split}
\vec{v}=v(t)\left(\sin kz , \cos kz, 0\right),
\end{split}
\end{equation}
where $A(t)$ and $v(t)$ are the amplitudes of $\vec{A}_{Y}$ and $\vec{v}$, respectively. In the following we also assume the presence of vorticity fluctuations in the plasma \cite{s3}. 
After neglecting the displacement current in the lab frame and using the aforementioned configurations, the hyperelectric field becomes \cite{Giovannini2000,Giovannini-2013oga,Giovannini-2015aea}  
\begin{equation}\label{eq16}
\vec{E}_{Y}=\frac{k^{\prime}}{\sigma }\vec{B}_{Y}-\frac{c_{\mathrm{v}}}{\sigma }k^{\prime}\vec{v}-\frac{c_{\mathrm{B}}}{\sigma}\vec{B}_{Y},
\end{equation}
and the evolution equation for the hypermagnetic field is obtained as
\begin{equation}\label{eq17}
\begin{split}
\frac{dB_{Y}(t)}{dt}=\left[-\frac{1}{ t}-\frac{{k^{\prime}}^{2}}{\sigma } +\frac{c_{\mathrm{B}}k^{\prime}}{\sigma } \right]B_{Y}(t)+\frac{c_{\mathrm{v}}}{\sigma}{k^{\prime}}^{2}v(t),
\end{split}
\end{equation}
where $\vec{\omega}=k^{\prime}\vec{v}$, $\sigma=100T$, and $k^{\prime}=k/R=kT$.
With the choice of vector potentials given in Eqs.\ (\ref{eq10},\ref{eq11}), $\langle\vec{v}(t).\hat{B}_{Y}(t)\rangle= v(t)$.\\

\section{Evolution equations for continuous spectra}\label{app-c}
In this appendix, we present the evolution equations for matter-antimatter asymmetries and hypermagnetic fields by generalizing the Chern-Simons configuration to continuum spectra. We decompose the vector fields into divergence-free eigenmodes of the Laplacian operator, $\vec{Q}^{\pm}(\vec{k})$, defined as \cite{s1}:
\begin{equation}
	\vec{Q}^{\pm}(\vec{k}) = \frac{\vec{e}_1(\vec{k}) \pm i\vec{e}_2(\vec{k})}{\sqrt{2}} \exp(i\vec{k} \cdot \vec{r}),
\end{equation}
where $\vec{e}_3 = \vec{k}/k$ and $(\vec{e}_1, \vec{e}_2, \vec{e}_3)$ are helicity basis that form a right-handed, orthonormal triad of unit vectors. 
These modes satisfy $\vec\nabla \cdot \vec{Q}^{\pm} = 0$ and $\vec\nabla \times \vec{Q}^{\pm} = \pm k \vec{Q}^{\pm}$, with $\vec{Q}^{\pm*}(-\vec{k}) = \vec{Q}^{\pm}(\vec{k})$.
For an incompressible fluid ($\vec\nabla \cdot \vec{v} = 0$), the velocity $\vec{v}$ and hypermagnetic field $\vec{B}$ are decomposed as \cite{s1}:
\begin{equation}
	\vec{v}(x, \vec{r}) = \int \frac{d^3k}{(2\pi)^3} \left[ \tilde{v}^+(x, \vec{k}) \vec{Q}^+(\vec{k}) + \tilde{v}^-(x, \vec{k}) \vec{Q}^-(\vec{k}) \right],
\end{equation}

\begin{equation}
	\vec{B}_Y(x, \vec{r}) = \int \frac{d^3k}{(2\pi)^3} \left[ \tilde{B}^+(x, \vec{k}) \vec{Q}^+(\vec{k}) + \tilde{B}^{-}(x, \vec{k}) \vec{Q}^-(\vec{k}) \right].
\end{equation}
where $ \tilde{v}^{\pm}(x, \vec{k}),\tilde{B}^{\pm}(x, \vec{k})$ are Fourier transforms of velocity and magnetic fields with positive and negative helicities, respectively. In this basis, the ensemble-averaged hypermagnetic field energy and helicity densities are:
\begin{equation}
	\frac{1}{2} \langle |\vec{B}_Y(x, \vec{r})|^2 \rangle \equiv \int d k \, E_B(x, k) = \int \frac{k^2 \, dk}{(2\pi)^2} \left[ |B^+(x, k)|^2 + |B^-(x, k)|^2 \right], 
\end{equation}

\begin{equation}\label{helicity}
	\langle \vec{A}_Y \cdot \vec{B}_Y \rangle \equiv \int d k \, H_B(x, k) = \int \frac{k \, dk}{2\pi^2} \left[ |B^+(x, k)|^2 - |B^-(x, k)|^2 \right],
\end{equation}
where, we have used statistically isotropic correlators,
\begin{equation}
	\langle \tilde{B}^{\pm*}(x, \vec{k}) \tilde{B}^{\pm}(x, \vec{k'}) \rangle = |B^{\pm}(x, k)|^2 (2\pi)^3 \delta^{(3)}(\vec{k} - \vec{k'}), \quad 
\end{equation}
\begin{equation}
	\langle \tilde{B}^{+*}(x, \vec{k}) \tilde{B}^{-}(x, \vec{k'}) \rangle = \langle  \tilde{B}^{-*}(x, \vec{k})  \tilde{B}^{+}(x, \vec{k'}) \rangle = 0. \quad 
\end{equation}

The fluid kinetic energy density is:
\begin{equation}
	\frac{\rho_r}{2} \langle |\vec{v}(x, \vec{r})|^2 \rangle \equiv \frac{\rho_r}{2} \int d k \, E_V(x, k) = \rho_r \int \frac{k^2 \, dk}{(2\pi)^2} \left[ |v^+(x, k)|^2 + |v^-(x, k)|^2 \right],
\end{equation}
where $\rho_r$ is the fluid density, and the velocity correlators satisfy:
\begin{equation}
	\langle \tilde{v}^{\pm*}(x, \vec{k}) \tilde{v}^{\pm}(x, \vec{k'}) \rangle = |v^{\pm}(x, k)|^2 (2\pi)^3 \delta^{(3)}(\vec{k} - \vec{k'}),
\end{equation}
\begin{equation}
	\langle \tilde{v}^{+*}(x, \vec{k}) \tilde{v}^{-}(x,  \vec{k'}) \rangle = \langle \tilde{v}^{-*}(x, \vec{k}) \tilde{v}^{+}(x, \vec{k'}) \rangle = 0.
\end{equation}
The hypermagnetic field evolution equation can be decomposed into equations for the modes $\tilde{B}^{\pm}$:
\begin{equation}\label{eq49-1} 
	\begin{split}
		&\frac{d}{dx}\tilde{B}^{+}(x, \vec{k})=\frac{1}{\sqrt{x}}\left[-C_{5}\left(\frac{k}{10^{-7}}\right)^{2}+ C_{6}\left(\frac{k}{10^{-7}}\right)\eta_{T}(x)
		-\frac{1}{\sqrt{x}}\right]\tilde{B}^{+}(x, \vec{k})\\&+\left(\frac{k}{10^{-7}}\right)^{2}\left[C_{7}\beta(x)+C_{8}\Delta \eta^{2}(x)\right]\frac{\tilde{v}^{+}(x, \vec{k})}{x^{3/2}},\\&	\frac{d}{dx}\tilde{B}^{-}(x, \vec{k})=\frac{1}{\sqrt{x}}\left[-C_{5}\left(\frac{k}{10^{-7}}\right)^{2}- C_{6}\left(\frac{k}{10^{-7}}\right)\eta_{T}(x)
		-\frac{1}{\sqrt{x}}\right]\tilde{B}^{-}(x, \vec{k})\\&+\left(\frac{k}{10^{-7}}\right)^{2}\left[C_{7}\beta(x)+C_{8}\Delta \eta^{2}(x)\right]\frac{\tilde{v}^{-}(x, \vec{k})}{x^{3/2}},		
	\end{split}
\end{equation}
where the coefficients $C_{i}, i=5,...,8$ are given in Eq.\ (\ref{eq51}). We multiply the first equation by $\tilde{B}^{+*}$ and the second by $\tilde{B}^{-*}$ and take ensemble averages to get

\begin{equation}\label{eq49-2} 
	\begin{split}
		&\frac{d}{dx}|B^+(x,k)|^2=\frac{2}{\sqrt{x}}\left[-C_{5}\left(\frac{k}{10^{-7}}\right)^{2}+ C_{6}\left(\frac{k}{10^{-7}}\right)\eta_{T}(x)
		-\frac{1}{\sqrt{x}}\right]|B^+(x,k)|^2\\&+2\left(\frac{k}{10^{-7}}\right)^{2}\left[C_{7}\beta(x)+C_{8}\Delta \eta^{2}(x)\right]\frac{\langle \tilde{v}^+(x,k)\tilde{B}^+(x,k)\rangle}{x^{3/2}}\\&
		\frac{d}{dx}|B^-(x,k)|^2=\frac{2}{\sqrt{x}}\left[-C_{5}\left(\frac{k}{10^{-7}}\right)^{2}- C_{6}\left(\frac{k}{10^{-7}}\right)\eta_{T}(x)
		-\frac{1}{\sqrt{x}}\right]|B^-(x,k)|^2\\&+2\left(\frac{k}{10^{-7}}\right)^{2}\left[C_{7}\beta(x)+C_{8}\Delta \eta^{2}(x)\right]\frac{\langle \tilde{v}^-(x,k)\tilde{B}^-(x,k)\rangle}{x^{3/2}},
	\end{split}
\end{equation}
We assume the following statistically isotropic correlators between velocity and hypermagnetic fields
\begin{equation}
	\begin{split}
		&\langle \tilde{B}^{\pm*}(x, \vec{k}) \tilde{v}^{\pm}(x, \vec{k'}) \rangle = |B^{\pm}(x, k)v^{\pm}(x, k)| (2\pi)^3 \delta^{(3)}(\vec{k} - \vec{k'}),\\&	
		\langle \tilde{B}^{+*}(x, \vec{k}) \tilde{v}^{-}(x, \vec{k'}) \rangle = \langle  \tilde{B}^{-*}(x, \vec{k})  \tilde{v}^{+}(x, \vec{k'}) \rangle = 0. \quad 
	\end{split}
\end{equation}
Assuming fully helical fields ($v^- = B^- = 0$), we define $M(x,k) \equiv |B^+(x,k)|^2$ and $F(x,k) \equiv \langle \tilde{v}^+ \tilde{B}^+ \rangle$. This leads to the following equation:

\begin{equation}\label{eq49a} 
	\begin{split}
		\frac{d}{dx}M(x,k)=&\frac{2}{\sqrt{x}}\left[-C_{5}\left(\frac{k}{10^{-7}}\right)^{2}+ C_{6}\left(\frac{k}{10^{-7}}\right)\eta_{T}(x)
		-\frac{1}{\sqrt{x}}\right]M(x,k)\\&+2\left(\frac{k}{10^{-7}}\right)^{2}\left[C_{7}\beta(x)+C_{8}\Delta \eta^{2}(x)\right]\frac{F(x,k)}{x^{3/2}} ,
	\end{split}
\end{equation}

\begin{equation}\label{eq49b} 
	\begin{split}
		\frac{d }{dx}F(x,k)=&\frac{1}{\sqrt{x}}\left[-C_{5}\left(\frac{k}{10^{-7}}\right)^{2}+ C_{6}\left(\frac{k}{10^{-7}}\right)\eta_{T}(x)
		-\frac{1}{\sqrt{x}}\right]F(x,k)\\&+\left[C_{7}\beta(x)+C_{8}\Delta \eta^{2}(x)\right]\left(\frac{k}{10^{-7}}\right)^{2}\frac{v(x,k)^2}{x^{3/2}} ,
	\end{split}
\end{equation}

Using the anomaly equation (\ref{er}) and Eq.\ (\ref{helicity}), the right-handed lepton asymmetries evolve as:

\begin{equation}\label{eq49c}
	\frac{d\eta_{\ell_R}}{dx} = -C \frac{d}{dx}\left[x^2 \int_{k_{\text{min}}}^{k_{\text{max}}} dk \, k M(x,k)\right] - \frac{\Gamma_\ell^0}{\sqrt{x}} \eta_{\ell,\text{Yuk}}(x),
\end{equation}
for $\ell = e, \mu, \tau$, where $C = 10^{-47} \frac{\alpha_Y}{10\pi^3 (106.75/45)}$.

The fluid kinetic energy spectrum is given by a power law \cite{s1}
\begin{equation}\label{ww2}
	E_V(x, k) \propto  k^n,
\end{equation}
On small length scales ($k > k_{\star}$), the Kolmogorov spectrum yields $n = -5/3$, while on large length scales ($k < k_{\star}$), we use a  Batchelor type spectrum with $n = 4$. Note that the initial spectrum for kinetic energy will generate the initial seed spectrum for the hypermagnetic energy and helicity spectra. The normalization constant is determined using Eq. (\ref{ww2}) \cite{Dvornikov:2022gkf}.
\begin{equation}\label{CKCB}
	\begin{split}
	\frac{v(x_0)^2}{2} &=\frac{1}{2} \int_0^{k_{\text{max}}} E_{V}(x_0,k) \, dk \\&=\frac{1}{(2\pi)^2} \int_0^{k_{\text{max}}} k^2|v^+(x_0,k)|^2 \, dk = C_B \int_0^{k_\star} k^{n_B} \, dk + C_K \int_{k_\star}^{k_{\rm max}}k^{n_K} \, dk , 
\end{split}
\end{equation}
where $v(x_0)$ is the initial seed of velocity and $k_{\star}=k_{\rm min}/\gamma^{\star}$ with $k_{\rm min}=10^{-14}$ and parameter $\gamma^{\star}$ lies in the range $10^{-3}<\gamma^{\star}<10^{-2}$ \cite{Dvornikov:2022gkf}. Moreover, we require the seed spectrum to be continuous, i.e., $C_B k_\star^{n_B} = C_K k_\star^{n_K}$. These conditions define constants $C_{B,K}$. Thus, the seed spectrum is given by \cite{Dvornikov:2022gkf}
\begin{equation}
	E_{V}(x_0,k) =v(x_0)^2 (1 + n_K) \left[ \frac{n_K - n_B}{1 + n_B} + \frac{ k_{\rm max}}{k_\star^{1 + n_K}} \right]^{-1} \times
	\begin{cases}
		\frac{k^{n_B}}{k_\star^{1 + n_B}}, & 0 < k < k_\star, \\
		\frac{k^{n_K}}{k_\star^{1 + n_K}}, & k_\star < k < k_{\rm max}.
	\end{cases}
\end{equation}
Therefore, $v^+(x_0,k)$ can be expressed as
\begin{equation}
	v^+(x_0,k)=\frac{\sqrt{2}\pi}{k}E_{V}(x_0,k)^{1/2}.
\end{equation}

%####################################################################################

{\bf Acknowledgments:} S. A. acknowledges the support of the Iran National Science Foundation (INSF) (grant No.\ 4003903). S. A. also acknowledges support by the European Union’s Framework Programme for Research and Innovation Horizon 2020
under the Marie Sklodowska-Curie grant agreement No 860881-HIDDeN as well as under
the Marie Sklodowska-Curie Staff Exchange grant agreement No 101086085-ASYMMETRY.

%(******************************************************************************************)
%\newpage
%\begin{multicols}{2}

%\end{multicols}
\end{document}